\documentclass[12pt]{article}
\pdfoutput=1
\usepackage[latin1]{inputenc}

\usepackage{amsmath}
\usepackage{amsfonts}
\usepackage{amssymb}
\usepackage{graphicx}
\usepackage{geometry}
\usepackage{amssymb,epsfig}
\usepackage{hyperref}
\usepackage{subfig}

\makeatletter
\renewcommand\section{\@startsection {section}{1}{\z@}%
                                 {-3.5ex \@plus -1ex \@minus -.2ex}
                                   {2.3ex \@plus.2ex}%
                                   {\normalfont\large\bfseries}}
\renewcommand\subsection{\@startsection{subsection}{2}{\z@}%
                                   {-3.25ex\@plus -1ex \@minus -.2ex}%
                                     {1.5ex \@plus .2ex}%
                                     {\normalfont\bfseries}}
\renewcommand\subsubsection{\@startsection{subsubsection}{3}{\z@}%
                                   {-3.25ex\@plus -1ex \@minus -.2ex}%
                                     {1.5ex \@plus .2ex}%
                                     {\normalfont\itshape}}
\makeatother

\def\pplogo{\vbox{\kern-\headheight\kern -29pt
\halign{##&##\hfil\cr&{\ppnumber}\cr\rule{0pt}{2.5ex}&\ppdate\cr}}}
\makeatletter
\def\ps@firstpage{\ps@empty \def\@oddhead{\hss\pplogo}%
  \let\@evenhead\@oddhead 
}
\def\maketitle{\par
 \begingroup
 \def\thefootnote{\fnsymbol{footnote}}
 \def\@makefnmark{\hbox{$^{\@thefnmark}$\hss}}
 \if@twocolumn
 \twocolumn[\@maketitle]
 \else \newpage
 \global\@topnum\z@ \@maketitle \fi\thispagestyle{firstpage}\@thanks
 \endgroup
 \setcounter{footnote}{0}
 \let\maketitle\relax
 \let\@maketitle\relax
 \gdef\@thanks{}\gdef\@author{}\gdef\@title{}\let\thanks\relax}
\makeatother

\numberwithin{equation}{section}

\renewcommand{\dag}{\dagger}

\newcommand{\be}{\begin{equation}}
\newcommand{\bea}{\begin{eqnarray}}
\newcommand{\ee}{\end{equation}}
\newcommand{\eea}{\end{eqnarray}}
\newcommand\beq{\begin{equation}}
\newcommand\eeq{\end{equation}}

\newcommand{\mc}{\mathcal}

\renewcommand{\t}{\tilde}

\newcommand{\ba}{\begin{align}}
\newcommand{\ea}{\end{align}}
\newcommand{\bg}{\begin{gather}}
\newcommand{\eg}{\end{gather}}
\newcommand{\bseq}{\begin{subequations}}
\newcommand{\eseq}{\end{subequations}}

\begin{document}

\setcounter{page}0
\def\ppnumber{\vbox{\baselineskip14pt
}}
\def\ppdate{\footnotesize{SLAC-PUB-14882 ~~~ SU-ITP-12/09}} \date{}

\author{Xi Dong, Bart Horn, Eva Silverstein, Gonzalo Torroba\\
[7mm]
{\normalsize \it Stanford Institute for Theoretical Physics }\\
{\normalsize  \it Department of Physics, Stanford University}\\
{\normalsize \it Stanford, CA 94305, USA}\\
[7mm]
{\normalsize \it Theory Group, SLAC National Accelerator Laboratory}\\
{\normalsize \it Menlo Park, CA 94309, USA}\\
}

\bigskip
\title{\bf  Unitarity bounds and RG flows in time dependent quantum field theory
\vskip 0.5cm}
\maketitle

\begin{abstract}
We generalize unitarity bounds on operator dimensions in conformal field theory to field theories with spacetime dependent couplings.  Below the energy scale of spacetime variation of the couplings, their evolution can strongly affect the physics, effectively shifting the infrared operator scaling and unitarity bounds determined from correlation functions in the theory.  We analyze this explicitly for large-$N$ double-trace flows, and connect these to UV complete field theories.     
One motivating class of examples comes from our previous work on FRW holography, where this effect explains the
range of flavors allowed in the dual, time dependent, field theory.

\end{abstract}
\bigskip
\newpage

\tableofcontents

\vskip 1cm




\noindent

\vspace{0.5cm}  \hrule
\def\thefootnote{\arabic{footnote}}
\setcounter{footnote}{0}

\section{Motivations}

Unitarity is essential to any closed quantum mechanical system such as Lorentzian-signature quantum field theory.
In conformal field theory, operator dimensions are bounded in a way that derives from unitarity.  For example for scalar operators ${\cal O}$ in a $d$-dimensional theory, the condition
\beq\label{scalarubound}
\Delta_{\cal O}\ge (d-2)/2
\eeq
follows from the positivity of the norm of states created by $\cal O$ and its descendants.  If one couples additional fields $\chi$ to the CFT via couplings of the form $\int d^dx g\chi {\cal O}$, the same condition (\ref{scalarubound}) arises from the optical theorem for forward scattering of $\chi$ (see~\cite{Grinstein:2008qk} for a clear recent discussion and~\cite{CFT} for general results on constraints from conformal invariance).

One reason unitarity bounds are interesting is that they can help determine aspects of the infrared physics of nontrivial quantum field theories.  For example, in supersymmetric QCD below the Seiberg window of conformal field theories, one can use this unitarity bound along with other symmetries to show that the theory cannot flow to a CFT~\cite{Seiberg:1994pq}.  In other examples, such as \cite{Barnes:2004jj,Kutasov:2003iy}, unitarity bounds help to establish that a certain field decouples from the rest of the theory in the infrared.

In this paper, we are concerned with the question of what happens to the infrared physics and unitarity bounds in field theories which have a time dependent coupling $g(t)$.\footnote{Works on spacetime dependent couplings and RG flows in QFT include~\cite{osborn,Komargodski:2011vj,disorder}.}  More specifically, we will study the infrared limit of field theories with spacetime dependent couplings that approximately preserve scale invariance but not Poincar\'e invariance over a wide range of spacetime scales.
Consider a coupling of the form
\beq\label{gterm}
\int d^dx {\cal L}_g=\int dt d^{d-1}\vec x \,g(t,\vec x){\cal O},
\eeq
with $g(t)\sim t^\alpha$ or $g(t,\vec x)\sim (t^2-\vec x^2)^{\alpha/2}$, at sufficiently late times, or sufficiently well within the forward light cone, respectively.  As we rescale the coordinates $x^\mu\to\lambda x^\mu$, the coupling transforms by the rescaling
\beq\label{gscale}
g(x)\to \lambda^{\alpha} g(x).
\eeq
If we perturb a CFT by $\int d^dx {\cal L}_g$ (\ref{gterm}), the nontrivial scaling of $g$ (\ref{gscale}) affects the question of whether ${\cal L}_g$ dominates at late times and large distance scales.  If ${\cal L}_g$ does dominate in the infrared and become marginal under our scaling by $\lambda$, then the nontrivial scaling of $g$ also affects the scaling of the operator ${\cal O}$.  This suggests that unitarity conditions and scaling dimensions of the infrared theory may be modified by the presence of $g(x)$.  Another way to organize this question, which we will consider below as well, is to realize the spacetime dependent coupling dynamically, as the profile of an additional field $g(x)$ in the theory.

To answer our question, we will analyze the infrared behavior of correlation functions in simple, tractable theories with such a coupling $g(x)$.  Clearly, the effects of the spacetime dependence of the coupling become unimportant on short enough scales, schematically $\Delta x\ll g/\partial g$.  For this reason, we focus on an infrared, late-time regime which we define carefully below. At shorter distance scales, we cross over to static physics.  We will consider examples below in which this static theory is well defined up to energy scales much larger than the scale $\partial g/g$, including a class of fully UV complete examples.

The general motivations to study time dependent quantum field theories are very simple.  Time dependent backgrounds are generic, and have important applications in fields as diverse as cosmology\footnote{(ultimately providing a simple theory of the origin of structure in the universe)} and condensed matter physics.\footnote{For example, quenching and thermalization is an interesting probe of field theories at finite density \cite{ADSCMT}. Nonequilibrium processes are also useful for stimulating interesting low temperature phases \cite{stimulate}.}  On a more theoretical front, it remains a major challenge to derive a framework for cosmology including the effects of horizons, singularities, and transitions among metastable configurations; various interesting approaches are being pursued.\footnote{Some recent approaches can be found in 
\cite{Bousso:1999dw, dSCFT, DioTom, Maldacena:2002vr, dSdS,Dong:2010pm, Dong:2011uf, Banks:2004vg, Freivogel:2006xu, Harlow:2010my, Harlow:2011az, McFadden:2009fg, Hertog:2011ky}. Other recent work on this general subject includes, for instance,~\cite{timecosmo1,DBI, Ross:2004cb, inside}.}      
Recently, by uplifting AdS/CFT solutions to cosmological ones (de Sitter and other FRW solutions), we made some progress on this question, and encountered the above issue.  Our uplifting procedure~\cite{Polchinski:2009ch, Dong:2010pm,Dong:2011uf} introduces a number $n$ of $(p,q)$ 7-branes, which introduce magnetic flavors into the dual theory.  For $n$ greater than a certain integer $n_*$ (depending on the dimensionality and other details of the example), one finds no static AdS/CFT solution. However, for $n>n_*$ one does find a controlled time dependent solution, which furthermore admits a warped metric indicating a low energy field theory dual.\footnote{A different class of solutions was analyzed in~\cite{Kleban:2007kk}.}  We analyzed the basic properties of this putative dual in several different ways, achieving consistent results. As we will show below by using the most supersymmetric class of examples, the problem with would-be static theories in the range $n>n_*$ can be traced to unitarity.  Our gravity solutions suggest that the time dependent background relaxes this condition.  Moreover, in our FRW solutions, the dual effective field theory is cut off at an energy scale proportional to $\partial g/g \sim 1/t$.

This led us to the above (more general) question, which we will address in much simpler examples than those of \cite{Dong:2011uf}.  In particular, we will focus on a theory with a spacetime dependent semi-holographic \cite{SemiHol}\ coupling $g(x)$ between a large-$N$ conformal field theory and a scalar field, including the simplest case corresponding to double trace renormalization group flows \cite{DoubleTraceorig}\cite{DoubleTraceflows}.  This system and its infrared unitarity for constant coupling was analyzed recently in \cite{Andrade:2011aa}\ in the case where the large-$N$ CFT has a large-radius holographic dual.  Here, we determine the effects of the spacetime dependence of $g(x)$ on the relevance of the coupling and the analysis of unitarity in the infrared theory.
In particular, we will find -- as anticipated in the FRW example just mentioned -- that spacetime dependent couplings can produce new interacting theories with approximate scale invariance in the infrared for a wider range of flavors than in the corresponding static theory.
Our results about the infrared physics will apply for parametrically long times at large $N$.\footnote{This is reminiscent of other large-$N$ field theories for which scaling symmetry holds over a wide range of scales but not arbitrarily far into the infrared.}

This paper is organized as follows.  In \S \ref{sec:semiholographic}, we present our main example, a semi-holographic model in which we can compute the needed amplitudes explicitly. In \S \ref{sec:fermion}, we analyze spacetime dependent couplings involving fermions, finding similar results. Then in \S \ref{sec:SQCD}, we consider supersymmetric gauge theories with time dependent couplings;  this provides a concrete UV completion of a class of models like those of \S \ref{sec:semiholographic}, and makes contact with our motivations from FRW holography.    Finally, we conclude in \S \ref{sec:concl} with additional comments and potential generalizations of our results. Several useful results are explained in more detail in the Appendix.

\section{Spacetime dependent double trace flows and semiholographic models}\label{sec:semiholographic}

Let us now analyze in detail our main example. We will study the RG evolution of a large-$N$ CFT in $d$ dimensions perturbed by a spacetime dependent coupling to a scalar field $\phi$:\footnote{Throughout, we will use mostly plus conventions for the metric signature. Our QFT conventions are those of Weinberg~\cite{Weinberg:1995mt}.}
\be\label{eq:S1}
S= S_{CFT} - \frac{1}{2} \int d^dx\, (\eta^{\mu\nu} \partial_\mu \phi \partial_\nu \phi + m^2 \phi^2) + \int d^dx\,g(x) \mc O \phi\,,
\ee
where $\mc O$ is an operator in the CFT. A related question that will be addressed is what happens when a CFT is deformed by a space-time dependent double-trace operator,
\be\label{eq:S2}
S = S_{CFT} + \int d^dx\,\lambda(x) \mc O^2\,.
\ee
These two systems are closely related: when the coupling $g$ in (\ref{eq:S1}) is relevant, integrating out a massive $\phi$ produces a double-trace deformation with $\lambda = g^2/(2m^2)$.

For many purposes, it is useful to think of $g(x)$ as a dynamical field rolling in a potential. Then questions about unitarity and backreaction from interactions and particle production can be understood more directly in the theory where $g$ is a dynamical field. This will be explored in detail in Appendix \S \ref{app:dynamicalg}, while in the rest of this section $g(x)$ will be treated as an external coupling. 

While in general it is a hard problem to determine the RG evolution of a QFT with spacetime dependent couplings, we will use the fact that in the large-$N$ limit the infrared dynamics can be calculated explicitly. We will first recall what happens in the static case, and then incorporate the effects of spacetime dependence. Various results and calculations are relegated to Appendices \S \ref{app:gaussian} and \S \ref{app:exp}.

\subsection{RG flow in the static limit}

Let us begin with the static case. The Lagrangian including the coupling between the CFT and $\phi$ is
\be
L= L_{CFT} - \frac{1}{2}  (\eta^{\mu\nu} \partial_\mu \phi \partial_\nu \phi + m^2 \phi^2) +g\, \mc O_\pm \phi
\ee
with $\mc O_\pm$ a CFT operator of dimension
\be
\Delta_{\pm} = \frac{d}{2} \pm \nu\;\;,\;\;\nu>0\,,
\ee
at the unperturbed $g=0$ fixed point, where $[g]=1 \mp \nu$. The two-point function is normalized as
\be\label{eq:O2ptspace}
\langle \mc O_{\pm}(x) \mc O_{\pm}(y) \rangle = \frac{1}{[(x-y)^2]^{\Delta_{\pm}}}\,. 
\ee
We analyze, in turn, three possibilities: coupling $\mc O_-$ or $\mc O_+$ to $\phi$ when $0<\nu<1$, or coupling $\mc O_+$ to $\phi$ for $\nu >1$.

First, consider the effect of coupling $\mc O_-$ to $\phi$. Unitarity of $\mc O_-$ requires $\nu<1$, and the coupling $g$ is relevant. At low energies the kinetic term for $\phi$ is negligible; integrating out $\phi$ sets
\be\label{eq:phiO}
\phi = \frac{g}{m^2} \mc O_-\,.
\ee
This produces a double-trace term $\frac{g^2}{2m^2} \mc O_-^2 \subset L$,
which is a relevant deformation of dimension $2 \nu$. It is known that this triggers a flow from $\mc O_-$ to $\mc O_+$, which we now review.

In order to calculate the dimension $\Delta(\mc O_-)$ in the infrared, it is convenient to first compute the two-point function for $\phi$ and then use (\ref{eq:phiO}) to obtain the correlator for $\mc O_-$. At large $N$, loops containing $\phi$ are negligible, and the $\phi$ two-point function is given by a geometric series which sums to
\be\label{eq:Om2pt}
\langle \phi(p) \phi(-p) \rangle = \frac{-i}{p^2 +m^2 - g^2 c_{-\nu} (p^2)^{-\nu}}\,.
\ee
This calculation is derived using path integrals in Appendix \S \ref{app:gaussian}. In particular, it uses the Fourier transform of the unperturbed correlator (\ref{eq:O2ptspace}) for $\mc O$,
\be\label{eq:int1}
\langle \mc O_{\pm}(p) \mc O_{\pm}(-p) \rangle = -i \,c_{\pm\nu} \,(p^2-i\epsilon)^{\pm\nu}\;,\;c_\nu\equiv 2^{-2\nu}\pi^{d/2}\frac{\Gamma(-\nu)}{\Gamma(\frac d2+\nu)}\,.
\ee
(Here $p^2 =-(p^0)^2 +(p^i)^2$, and in what follows the $i \epsilon$ prescription will be implicit in our formulas.) The last term in the denominator of (\ref{eq:Om2pt}) dominates at low energies,
\be
\langle \phi(p) \phi(-p) \rangle  \approx i \frac{(p^2)^\nu}{g^2 c_{-\nu}}\,,
\ee
corresponding to an operator of dimension $\Delta_+$. Thus, $\Delta_{IR}(\mc O_-) = \Delta_+$, and we recover the double-trace flow from $\mc O_-$ to $\mc O_+$.

Next, let us instead couple $\mc O_+$ to $\phi$, still in the range $0< \nu <1$. The $g \mc O_+ \phi$ interaction is relevant, and the two-point function for $\phi$ becomes
\be\label{eq:Op}
\langle \phi(p) \phi(-p) \rangle = \frac{-i}{p^2 +m^2 - g^2 c_{\nu} (p^2)^{\nu}}\,.
\ee
The difference with the previous case is that now the $g^2$ contribution is less important than the mass term, so at low energies the correlator can be expanded in inverse powers of the mass,
\be\label{eq:Oprel}
\langle \phi(p) \phi(-p) \rangle \approx \frac{-i}{m^2}\left(1 + \frac{g^2}{m^2} c_\nu (p^2)^\nu +\ldots \right)\,,
\ee
corresponding to an operator of dimension $\Delta_+$.\footnote{The first term in (\ref{eq:Oprel}) gives a contact term that has to be subtracted when relating $\phi$ and $\mc O$ inside the path integral via $\phi =\frac{g}{m^2} \mc O$.} This implies that the dimension of $\mc O_+$ does not change in flowing to the IR. This may be understood by noting that the double trace deformation $\frac{g^2}{2m^2} \mc O_+^2 \subset L$ obtained by integrating out $\phi$ is actually an irrelevant perturbation of the $g=0$ conformal fixed point.

Finally, we come to the range $\nu >1$ and consider an interaction $g \mc O_+ \phi$ ($\mc O_-$ does not exist in this case since it would violate the unitarity bound). This was the static theory studied in~\cite{Andrade:2011aa}. The propagator for $\phi$ is still given by (\ref{eq:Op}) but, crucially, now $g$ is irrelevant. As a result, in the IR we simply have the original CFT plus a decoupled scalar field. Conversely, in the UV $g$ becomes strong at a scale of order
\be
\Lambda_g \approx \frac{1}{g^{\frac{1}{\nu-1}}}\,.
\ee
Choosing for simplicity a mass parameter $m \ll \Lambda_g$, for $\nu>1$ the propagator (\ref{eq:Op}) has a new pole at $p^2 \approx 1/(c_\nu g^2)^{1/(\nu-1)}$, which moreover has a residue with the opposite sign as the usual one at $p^2 \approx -m^2$. We thus learn that the theory has a tachyonic ghost and violates unitarity around the scale $\Lambda_g$. If we anyway continue past $\Lambda_g$ towards the UV, we may interpret the $\phi$ two-point function as giving an `inverse' RG flow to $\mc O_-$. Of course, the theory by itself is inconsistent and needs a UV modification above the scale $\Lambda_g$.

At this point it is useful to discuss a physical way of deriving unitarity bounds in the static theory proposed by~\cite{Grinstein:2008qk}, which we will then apply to the time dependent theory. The idea is to couple a probe scalar $\chi$ to the operator of interest $\mc O$ via $\chi \mc O \subset L$; then requiring that the $\chi \to \chi$ amplitude satisfy the optical theorem reproduces the unitarity bound for $\mc O$. In more detail, in terms of (\ref{eq:int1}),
\be\label{eq:scattering}
{\rm Im}\,\mc A(\chi \to \chi) \propto  {\rm Im}\,i \langle \mc O(p) \mc O(-p) \rangle  \propto - c_\nu\,\sin(\pi \nu)\,.
\ee
Then ${\rm Im}\,\mc A \geq 0$ for $\nu>-1$, which is the unitarity bound $\Delta_+ \geq \frac{d-2}{2}$ for $\mc O_+$.

To summarize, the interaction $g \phi \mc O$ can either give a flow from $\mc O_-$ to $\mc O_+$ in the IR, a double trace deformation $\mc O_+^2$ that is irrelevant at the $g=0$ fixed point with the scalar $\phi$ becoming strongly coupled in the IR, or a theory where $\phi$ decouples in the IR but its interaction with $\mc O_+$ is nonrenormalizable and violates unitarity around its strong coupling scale $\Lambda_g$ (above which we would formally get $\mc O_-$). In what follows we will study how all this is modified when $g$ becomes spacetime dependent.

\subsection{Spacetime dependent case}

Having reviewed the static limit, we now consider a spacetime dependent interaction
\beq\label{tdepmix}
g(x) \phi \mc O_+ \subset L
\eeq
and study the long distance dynamics. Since we are interested in the possibility of a new scale invariant regime at long distance, we will assume a power-law dependence approaching
\be\label{eq:gx}
g(x) \to g_0 (t^2- \vec x^2)^{\alpha/2}\;\;\; {\rm or} \;\;\;g(x)\to g_0|t|^\alpha \;,\;\alpha>0\,, 
\ee
well within the forward light cone, or at late times, respectively. With this in mind, in what follows we denote (\ref{eq:gx}) simply by $g(x)= g_0|x|^\alpha$. We will show in \S \ref{app:dynamicalg} that this dependence arises when a dynamical field $g$ rolls in a potential $V \propto -g^{2\left(1-\frac{1}{\alpha} \right)}$.\footnote{For even $\alpha$, it is possible choose a state defined by analytically continuing to the Euclidean theory with $g(x)= g_0|x|^\alpha$ and imposing regularity at $x^\mu=0$.}


Before getting into any detailed calculation, let us start simply by noting the scaling of the double trace interaction induced by our mixing term (\ref{tdepmix}), taking the rest of the Lagrangian to depend on $\phi$ simply as $m^2\phi^2$.  Integrating out $\phi$ leads as in the static case to a double trace deformation
\beq\label{dtracet}
\int d^dx\, \frac{g(x)^2}{m^2}{\cal O}_+^2 .
\eeq
With the spacetime dependent $g(x)$ in (\ref{eq:gx}), the effective coupling is $g_0^2/m^2$.  This has dimension
\beq\label{geffdim}
\left[\frac{g_0^2}{m^2} \right]=2(\alpha-\nu)\,.
\eeq
This indicates that the condition for relevance of (\ref{tdepmix}) is shifted by $\alpha$ relative to the static case, to $\alpha>\nu$.  As a result, we expect that even for cases with $\nu>0$ in which the double trace deformation would be irrelevant in the static case (including cases with $\nu>1$ in which the irrelevant deformation leads to a pathological theory in the UV), as long as $\alpha >\nu$ the term (\ref{tdepmix}) will be relevant.  Our calculations of correlation functions below will bear this out.\footnote{We will also generalize to include the possibility that $\phi$ itself also starts with a kinetic term $-\int d^dx (\partial\phi)^2$ in the UV.}

We are particularly interested in how (\ref{eq:gx}) modifies the dynamics in the range $\nu>1$, for which in the static theory unitarity violation arises in the UV and $\phi$ becomes a free field in the IR. It is important to point out that for $\nu>1$ the theory will still need a UV completion or cutoff, because at momenta much larger than the rate of change $\partial g/g$, the static limit is recovered. On the other hand, we will show that for momenta $\Delta p \ll (\partial g/g)$ the spacetime dependent interaction changes the theory in important ways, depending on the relation between $\alpha$ and $\nu$. Similar effects will be found when $0<\nu<1$, though in this range the theory is UV complete since the static theory is consistent at high energies.

The calculation of the two-point function of $\phi$ is similar to the static one (\ref{eq:Om2pt}), and can be done directly in position space for a general $g(x)$. Denoting the propagators of the unperturbed $g=0$ theory by
\bea\label{eq:kokp}
K_{\mc O_+}^{-1}(x,y)&\equiv& i \langle \mc O_+(x) \mc O_+(y) \rangle_{g=0}= c_\nu \int \frac{d^dp}{(2\pi)^d} e^{ip(x-y)}(p^2-i\epsilon)^\nu = \frac{i}{[(x-y)^2]^{\Delta_+}}\nonumber\\
K_\phi^{-1}(x,y)&\equiv& i \langle \phi(x) \phi(y) \rangle_{g=0}= \int \frac{d^dp}{(2\pi)^d} \frac{e^{ip(x-y)}}{p^2+m^2-i\epsilon}\,,
\eea
the two-point function for $\phi$ including quantum effects from $\mc O_+$ insertions but ignoring $\phi$ loops becomes
\be
\langle \phi(x) \phi(x')\rangle=-i K_\phi^{-1}(x,x') -i \int d^dz_1 d^dz_2 K_\phi^{-1}(x,z_1) g(z_1) K_{\mc O_+}^{-1}(z_1,z_2) g(z_2) K_\phi^{-1}(z_2,x')+ \ldots
\ee
This is again a geometric series, which in matrix notation sums to\footnote{Here the inverse $K^{-1}(x,y)$ means $\int_z K(x,z) K^{-1}(z,y)=\delta^d(x-y)$, and $g$ can be thought of as a diagonal matrix $g(x,y)=g(x)\delta^d(x-y)$. Appendix \S \ref{app:gaussian} presents a similar derivation in gaussian theories directly from the path integral.}
\be\label{eq:phiphi}
\langle \phi(x) \phi(x')\rangle=-i \left(K_\phi- g K_{\mc O_+}^{-1} g \right)^{-1}(x,x')\,.
\ee

Recall that the static derivation requires large $N$ at fixed $g$, so that internal loops containing $\phi$ are suppressed by powers of $1/N$. In the present spacetime dependent situation, we need to be more careful about this since our specified $g(x)$ grows at large $x$.   Ignoring $\phi$ loops as we just did is valid as long as $g(x) \ll N^\gamma$, with $\gamma \sim 1$ depending on the details of the CFT OPEs. Thus, at large $N$ and for a power-law interaction, the correlator (\ref{eq:phiphi}) starts receiving corrections at parametrically large distances $|x| \sim N^{\gamma/\alpha}$. In what follows we restrict to scales where such effects are negligible.

We thus find that at large $N$, quantum effects from the spacetime dependent interaction can be calculated explicitly in our semiholographic model, and they lead to the effective action for $\phi$
\be\label{eq:Seff}
S_\textrm{eff}[\phi]=- \frac{1}{2} \int d^dx\,d^dx'\, \phi(x) \left((-\partial_x^2+m^2)\delta^d(x-x')-\frac{i g(x) g(x')}{[(x-x')^2]^{\Delta_+}} \right)\phi(x') \,,
\ee
which is equivalent to (\ref{eq:phiphi}).  Note that the factor of $i$ appearing in the last term of the effective action does not necessarily violate unitarity, since this is a nonlocal term.  Our analysis so far has been for a general $g(x)$. Next we will specialize to (\ref{eq:gx}) and we will analyze how the decoupling of $\phi$ in the static limit is modified by the spacetime dependence in the action (\ref{eq:Seff}).

\subsection{Long distance propagator}

Let us now focus on momenta $\Delta p \ll (\partial g/g) \sim 1/|x|$ and determine the propagator when the last term in \eqref{eq:Seff} dominates. This will be valid in a certain range of $\nu$ and $\alpha$ that we find below by requiring the effects from $K_\phi(x,x')$ to be negligible.

Ignoring the first two terms in \eqref{eq:Seff} and keeping only the contributions proportional to $g(x)g(x')$, the effective action becomes
\be\label{eq:Seffapprox}
S_\textrm{eff}[\phi]= \frac{1}{2} \int d^dx\,d^dx'\, \frac{i \t \phi(x) \t \phi(x')}{[(x-x')^2]^{\Delta_+}} \,,
\ee
with $\t \phi(x) \equiv g(x) \phi(x)$. From this it is clear that the $\tilde \phi$ propagator is -- up to a sign which we will determine next -- that of ${\cal O}_-$.
In particular, Fourier transforming $\t \phi(x)$ we obtain
\be\label{eq:Seff2}
S_\textrm{eff}[\phi]= \frac{1}{2} \int \frac{d^dp}{(2\pi)^d}\, \tilde\phi(p)\tilde\phi(-p)\, c_\nu (p^2-i\epsilon)^\nu,
\ee
where again the coefficient
\be
c_\nu=2^{-2\nu}\pi^{d/2}\frac{\Gamma(-\nu)}{\Gamma(\frac d2+\nu)}\,.
\ee

From (\ref{eq:Seff2}), the two-point function for $\tilde\phi(p)$ is
\be\label{twotildep}
\langle\tilde\phi(p)\tilde\phi(-q)\rangle=i\delta^d(p-q)\,\frac{(p^2-i\epsilon)^{-\nu}}{c_\nu}\,.
\ee
We can then Fourier transform back to position space and get
\be\label{twotilde}
\langle\tilde\phi(x)\tilde\phi(x')\rangle =\frac{i}{c_\nu}\int \frac{d^dp}{(2\pi)^d}\, e^{ip\cdot(x-x')}(p^2-i\epsilon)^{-\nu}=\frac{-1}{c_\nu c_{-\nu}[(x-x')^2]^{\Delta_-}}\,.
\ee
Recalling the relation between $\t \phi(x)$ and $\phi(x)$, we arrive at our final expression for the two-point function
\be\label{eq:finalGphi}
\langle\phi(x)\phi(x')\rangle=\frac{-1}{c_\nu c_{-\nu}g(x)g(x')[(x-x')^2]^{\Delta_-}}\,,
\ee
up to corrections from the first two terms in (\ref{eq:Seff}).

We now recall that the massive case is equivalent to a spacetime dependent double trace deformation (\ref{dtracet}) of the original CFT. From the relation between $\phi$ and $\mc O_+$ and the result (\ref{eq:finalGphi}), we conclude that the double trace perturbation induces a flow for $\mc O_+$ between the UV two-point function (\ref{eq:O2ptspace}) and a long distance/late times correlator
\be\label{eq:OpIR}
\langle\mc O_+(x)\mc O_+(x')\rangle=\frac{-m^4}{c_\nu c_{-\nu}g(x)^2 g(x')^2[(x-x')^2]^{\Delta_-}}\,.
\ee
This is one of our main results, exhibiting the effects of the spacetime dependent coupling on the IR dynamics.
While translation invariance has been broken explicitly, the power-law dependence of this correlator signals the appearance of a new scale invariant regime. This two-point function implies that, under a dilatation $(x,x')\to\lambda (x,x')$, the operator transforms as $\mc O_+ \to \lambda^{-\left(\frac{d}{2}- \nu+ 2\alpha \right)} \mc O_+$. We should also stress that this regime is approximate, being modified by small nonadiabatic effects as well as by $1/N$ corrections.

\subsubsection{Relevance condition}

Before exploring the physics contained in this result, let us analyze its regime of applicability. In other words, we need to find out when the last term in \eqref{eq:Seff} is relevant and dominates in the IR. We first analyze this with a simple scaling argument which reproduces (\ref{geffdim})  in the case that the $\phi$ mass dominates, and then show the same result by expanding the two-point function.

Under the dilatation $(x,x')\to\lambda (x,x')$, the term $\partial_x^2\delta^d(x-x')$ transforms with weight $d+2$, $m^2\delta^d(x-x')$ transforms with weight $d$, and $g(x)g(x')/[(x-x')^2]^{\Delta_+}$ transforms with weight $2(\Delta_+-\alpha)$ in either the Lorentz invariant case $g(x)=g_0|x|^\alpha$ or the purely time-dependent case $g(x)=g_0|t|^\alpha$. Therefore the $g(x)g(x')$ term is more relevant than the other terms if we satisfy
\be\label{eq:rel}
\alpha>\begin{cases}\nu\,,&\quad m\ne0\\\nu-1\,,&\quad m=0\end{cases}
\ee

We may reproduce these conditions by expanding the full two-point function \eqref{eq:phiphi} around $K_\phi=0$:
\be
\langle\phi(x)\phi(x')\rangle =i \left[(g K_{\mc O_+}^{-1} g)^{-1}+(g K_{\mc O_+}^{-1} g)^{-1} K_\phi (g K_{\mc O_+}^{-1} g)^{-1}+\cdots \right](x,x')\,.
\ee
The leading term here is given by \eqref{eq:finalGphi}. The subleading term from $K_\phi$ can be neglected if
\be\label{eq:Kphismall}
\left|(g K_{\mc O_+}^{-1} g)^{-1} K_\phi (g K_{\mc O_+}^{-1} g)^{-1}(x,x')\right|\ll \left|(g K_{\mc O_+}^{-1} g)^{-1}(x,x')\right|\,.
\ee
We ask when this is true and the two-point function is well approximated by \eqref{eq:finalGphi} at long distance/late times (for large $|x|$, $|x'|$, and $|x-x'|$). We show in Appendix \S \ref{subsec:expn} that the condition is the same as \eqref{eq:rel}. These results are in agreement with the effective dimension (\ref{geffdim}) calculated in the UV.

We may also ask under what conditions the free fixed point $g=0$ is stable. In other words we expand the full two-point function \eqref{eq:phiphi} (and the one for $\mc O$) around $g=0$, and ask when the subleading terms are much smaller than the leading one. We show in Appendix \S \ref{subsec:expf} that this condition is precisely the opposite of \eqref{eq:rel}. This is what we expect: if \eqref{eq:rel} is satisfied, our theory is dominated in the IR by the last term in the effective action \eqref{eq:Seff}, otherwise it is dominated by the first two terms in \eqref{eq:Seff}, giving a free, decoupled scalar field.

When the coupling is a function of $t$ only, $g= g_0 |t|^\alpha$, it is more convenient to study the relevance of the different terms in the action by first Fourier transforming the spatial coordinates. This gives, in Euclidean space,
\be\label{eq:Smixed}
S_\textrm{eff}=- \frac{1}{2} \int_{t,t',\vec p}  \phi_{\vec p}(t) \left((\partial_t^2+m^2+ \vec p^{\,2})\delta(t-t')-C_\nu g(t) g(t') |\vec p|^{2\nu+1}\frac{K_{\nu+\frac{1}{2}}(|\vec p| |t-t'|)}{(|\vec p| |t-t'|)^{\nu+ \frac{1}{2}}} \right)\phi_{-\vec p}(t')
\ee
where $C_\nu$ is a positive constant from the Fourier transform of $1/[(t-t')^2 + \vec x^{\,2}]^{\Delta_+}$ with respect to $\vec x$. The two-point function in this mixed representation reads
\be
\langle \phi_{\vec p}(t) \phi_{-\vec p}(t') \rangle = \left(K_{\phi, \vec p}- g K^{-1}_{\mc O_+, \vec p} \,g \right)^{-1}(t,t')
\ee
where $K_{\phi, \vec p}$ and $g K^{-1}_{\mc O_+, \vec p} \,g$ refer to the first and second term in (\ref{eq:Smixed}). The condition that the $g K^{-1}_{\mc O_+, \vec p} \,g$ term dominates (which is still given by (\ref{eq:Kphismall}) but now in the mixed $(t, \vec p)$ basis) gives again $\alpha>\nu$ in the massive case and $\alpha>\nu-1$ in the limit $m=0$, in agreement with (\ref{eq:rel}).

\subsection{Infrared physics and unitarity: two wrongs make a right}\label{subsec:twowrongs}

We would now like to generalize and analyze standard unitarity conditions in our theories.
In ordinary quantum field theory, a theory that is well defined at a scale $\Lambda$ will retain unitarity in the infrared. With a good UV complete theory, in other words, it is not that unitarity bounds can fail; but they are useful in helping constrain the low energy behavior of the theory.  We expect that the same holds for systems like ours which are static at high energies but are subject to nontrivial time dependent effects at lower energies.

Above the scale $\partial g/g$, the coupling in our theory becomes effectively static.  We may UV complete the theory in various ways; in  \S \ref{sec:SQCD} we will describe a supersymmetric completion and in Appendix \S \ref{app:dynamicalg} we will describe a partial UV completion in which the time dependent coupling is the homogeneous mode of a dynamical field.  Without implementing such a procedure, however, starting at any given time we may tune the value of the coupling to obtain a separation of scales between the scale $\Lambda_g$ in which the static coupling becomes strong and the scale $\partial g/g$.  This hierarchy eventually breaks down at late times when $\alpha > \nu - 1$, but can be tuned to apply over a parametrically long region in time.\footnote{In some cases, it may happen that the theory is sensible even when the nominal scale $\Lambda_g$ goes below the scale $\partial g/g$, but we have not fully analyzed such cases.  A toy model in which we set up the corresponding question appears in Appendix \S \ref{app:higherderiv}.  A related question is whether one can implement a cutoff at the scale $\partial g/g$; this arises at the level of a radial cutoff in the holographic models of FRW cosmology in \cite{Dong:2011uf}.}  


Let us now turn to a direct analysis of unitarity conditions in our theories.
We have found that time dependence can significantly affect the infrared behavior of correlation functions.  It is particularly interesting to consider $\nu>1$. In this case, for constant coupling $g$ (i.e. $\alpha=0$), the semiholographic mixing term in the Lagrangian is irrelevant; the operator we are considering has dimension $\Delta_+$ in the infrared, rather than flowing to an operator of dimension $\Delta_-$ (which would violate the unitarity bound).
However, in the time dependent version of the $\nu>1$ theory, with $g(t)=g_0|t|^\alpha$, if we choose $\alpha$ appropriately we have seen that the semiholographic deformation can dominate in the infrared.
Moreover, at long distances this theory has a two-point function for $\tilde\phi$ (\ref{twotilde}) which scales like that of an operator of dimension $\Delta_-<(d-2)/2$.  In addition, from (\ref{twotilde}) we see that the two-point function in position space is negative for some ranges of $\nu$, including $1<\nu<2$.  In a conformal field theory, either of these features would imply a failure of unitarity.  The negative two-point function would correspond to a negative norm state created by $\tilde\phi$.  If the two-point function of $\tilde\phi$ were positive, the subunitary operator dimension $\Delta_-$ would imply that the descendant $\partial^2\tilde\phi$ creates a state of negative norm.\footnote{For a primary operator $\mathcal{O}$ in a CFT, the norm of the state created by the descendant $\partial^2 \mathcal{O}$ can be related to an expression containing the two-point functions $\langle\mathcal{O} \partial^2\mathcal{O}\rangle$ and $\langle\partial^{\mu}\mathcal{O} \partial^2\mathcal{O}\rangle$ as well as $\langle\partial^2\mathcal{O} \partial^2\mathcal{O}\rangle$, which are easily computed by taking derivatives of the propagator $1/|x-y|^{2\Delta_-}$.  The final result has a sign equal to ${\rm sgn}(\Delta_--(d-2)/2)$.}

In our case, on scales where the time dependent coupling affects the physics we lose the constraints of conformal symmetry, and we cannot compute the norm of states as in the static conformal field theory.  However, we can use the method discussed recently for conformal field theories in \cite{Grinstein:2008qk} to obtain a unitarity condition both in the static and time dependent theories. In that way of organizing the problem, we use the momentum space propagator (\ref{twotildep}) to compute the imaginary part of the forward scattering amplitude for Fourier modes, which must be positive for unitarity.  This quantity, as shown in \cite{Grinstein:2008qk}, has a sign equal to ${\rm sgn}[C(\Delta-(d-2)/2)]$ for an operator of dimension $\Delta$ with two-point function $C/|x-y|^{2\Delta}$.

We can now come to the key point:\ the imaginary part of the scattering amplitude  calculated using (\ref{twotildep}) has a sign equal to that of $-\sin(\pi\nu)/c_\nu$. See also the discussion around (\ref{eq:scattering}). This is positive for all $\nu$ (recall $\nu>0$), implying a positive imaginary part for the forward scattering amplitude.  This is consistent with the general expectation that all theories in the class we are discussing -- including both spacetime dependent ($\alpha\ne 0$) and static ($\alpha=0$) cases -- are unitary in the infrared.  

It is important for this argument that the sign of the position space propagator only determines the sign of the norm of a state in the static case.  Of course
on short scales where the time dependent couplings do not affect the physics, the theory reverts to a static one.  But on those scales, the correlation function is not of the form (\ref{twotildep}) (\ref{twotilde}), and the unitarity conditions are the familiar ones.  With a sensible completion of the theory above the scale $\partial g/g$, such as those mentioned at the beginning of this section, the static unitarity conditions are satisfied in this regime.

\section{Spacetime dependent RG flow for fermionic operators}\label{sec:fermion}

Our analysis so far has focused on scalar operators. Now we will couple a fermion $\psi$ to a fermionic operator $\mc O_f$ in a large-$N$ CFT via a spacetime dependent interaction\footnote{We employ the two-component notation of~\cite{twocomp}, which is well adapted for calculations in $d=4$. All the fermions are Weyl and transform as $(1/2,0)$; also, $\bar \psi_{\dot \alpha} \equiv (\psi_\alpha)^\dag$. In this section we will restrict our discussion to $d = 4$;\ however, the results below can be adapted to other dimensions by changing the representations of $\gamma$ matrices.}
\be\label{eq:Lfermion}
L = L_{CFT} - i \bar \psi \bar \sigma^\mu \partial_\mu \psi - \frac{1}{2} m \psi \psi+ g(x) \psi \mc O_f + c.c.
\ee
and study the RG evolution. In the massive case (\ref{eq:Lfermion}) leads to a spacetime dependent double trace perturbation of the original CFT,
\be\label{eq:doubletOf}
\int d^dx\, \frac{g(x)^2}{m} \mc O_f^2\,,
\ee
and we will determine its effects at long distances. The main conclusions are similar to the ones described in \S \ref{sec:semiholographic}: at large $N$ the system has an effectively gaussian description which, for large enough $\alpha$, flows to an approximately scale invariant regime controlled by the time dependent coupling. As before, this will also be consistent with scaling arguments around the UV fixed point. More explicit calculations for the fermionic gaussian model are relegated to Appendix \S \ref{app:gaussian}.

One motivation for looking at fermionic theories comes from supersymmetry. For instance, the UV completion that we discuss in \S \ref{sec:SQCD} is based on supersymmetric QCD and features an interaction between the fermionic part of the meson superfield and a fermionic operator made of magnetic quarks. We should nevertheless stress that our analysis below does not assume supersymmetry, and in fact we will find that the basic fermionic double trace flow is not related directly by supersymmetry to the previous bosonic double trace result. 
Another application of the fermionic version (which would be an interesting future direction), is to semiholographic Fermi surfaces with time dependent interactions. The static version of this model was proposed by~\cite{Faulkner:2010tq}  in $2+1$ dimensions. Static double-trace fermionic deformations have been studied recently in~\cite{Laia:2011wf,Allais:2010qq}; from the bulk perspective, the double-trace operator generated by (\ref{eq:doubletOf}) corresponds to a Majorana-type deformation in the classification of~\cite{Laia:2011wf}.

\subsection{Results for the static theory}

As in the bosonic case, we will see that the new IR regime dominated by the time dependent coupling has a simple description in terms of a static correlator for $\t \psi(x) \equiv g(x) \psi(x)$, so it is useful to first analyze the static version of the theory.

We find it convenient to denote the fermionic CFT operators that couple to $\psi$ by $\mc O_{f\pm}$, with dimensions at the $g=0$ fixed point parametrized by
\be
\Delta_{f \pm} = \frac{d}{2} \pm \left(\nu - \frac{1}{2} \right)\,.
\ee
Both operators are unitary for $0<\nu<1$, corresponding to the standard and alternate quantizations of fermionic fields in AdS/CFT.  Quantum corrections will again be given by a geometric series containing the $\mc O_f$ propagator,
\be\label{eq:OfOb}
\langle \mc O_f(x) \overline{\mc O}_f(y) \rangle = -i \sigma^\mu \frac{\partial}{\partial x^\mu} \,\frac{1}{[(x-y)^2]^{\Delta_f- \frac{1}{2}}}= (2\Delta_f-1)  \frac{i \sigma \cdot (x-y)}{[(x-y)^2]^{\Delta_f +\frac{1}{2}}}\,.
\ee
The dependence on the coordinates and $\sigma$ matrices is fixed by Lorentz and conformal invariance, and the normalization has been chosen to be consistent with our normalization (\ref{eq:kokp}) in cases when there is supersymmetry.\footnote{Eq.~(\ref{eq:OfOb}) is then a consequence of the identity
$$
\langle \mc O_f(x)_\alpha \delta_\xi \mc O^*_b(y) \rangle + \langle \delta_\xi \mc O_f(x)_\alpha \mc O^*_b(y) \rangle=0
$$
where the supersymmetry variations are given by $\delta_\xi \mc O_b = \xi \mc O_f\;,\;\delta_\xi \mc O_f = i \sigma^\mu \bar \xi\,\partial_\mu \mc O_b\,$. See e.g.~\cite{Dolan:2001tt} for properties of correlation functions in superconformal field theories.
}
We will also need the correlator in momentum space,
\be\label{eq:Of2ptp}
\langle \mc O_{f+}(p) \overline{\mc O}_{f+}(q) \rangle = i \sigma \cdot p\,c_{\nu-1} (p^2-i\epsilon)^{\nu-1}\,\delta^d(p-q)\,.
\ee
The result for $\mc O_{f-}$ follows by replacing $\nu-1 \to -\nu$.

Resumming the geometric series with quantum corrections from $\mc O_{f\pm}$ gives a self-energy for $\psi$ proportional to the propagator (\ref{eq:Of2ptp}). After inverting the Pauli matrices, the two-point function becomes
\be\label{eq:staticGpsi}
\langle \psi(p) \bar \psi(p) \rangle =-i \sigma \cdot p\, \frac{1-c_\gamma|g|^2 (p^2)^\gamma }{p^2\left(1-c_\gamma |g|^2 (p^2)^\gamma \right)^2 +m^2 - i \epsilon } 
\ee
where we have defined $\gamma \equiv \pm \left(\nu - \frac{1}{2} \right)-\frac{1}{2}$ for $\mc O_{f\pm}$ respectively. 

Let us now briefly discuss the different consequences of this result. We restrict to $\nu>1/2$ so that $\Delta_{f-}< \Delta_{f+}$; choosing $\nu<1/2$ just interchanges the roles of $\mc O_{f+}$ and $\mc O_{f-}$. First, when $\psi$ interacts with $\mc O_{f-}$ via $g \psi \mc O_{f-}\subset L$, the dimension of the interaction is $[g]=\nu$ and the double trace deformation $\mc O_{f-}^2$ has dimension $2\nu-1$, so both are relevant. We need to restrict to $\nu<1$ for unitarity. The long distance correlator becomes
\be\label{eq:OmfOpf}
\langle \psi(p) \bar \psi(p)\rangle \approx i \sigma \cdot p\, \frac{(p^2-i\epsilon)^{\nu-1}}{c_{-\nu}|g|^2}
\ee
corresponding to an operator of dimension $\Delta_{f+}$.\footnote{The correlator of a fermionic operator $\mc O_f$ of dimension $\Delta_f$ scales like $\langle \mc O_f(p) \overline{\mc O}_f(p) \rangle \sim i\sigma \cdot p\,(p^2)^{\Delta_f - \frac{d}{2}- \frac{1}{2}}$.} In the massive case this then describes a fermionic double trace flow from $\mc O_{f-}$ to $\mc O_{f+}$. This is the analog of the bosonic double trace flow between $\mc O_-$ and $\mc O_+$, although the fermionic version differs in the shifts by $1/2$. This difference has nontrivial consequences for supersymmetric theories. It means that there cannot be a supersymmetric double trace flow between superfields $\mc O_-$ in the UV and $\mc O_+$ in the IR. For instance, if we start with a supersymmetric $(\mc O_{b-}, \mc O_{f-})$ of dimensions $(\frac{d}{2}-\nu,\frac{d}{2}-\nu+ \frac{1}{2})$, the bosonic flow ends on a scalar operator of dimension $\frac{d}{2}+ \nu$ that would have a fermionic partner of dimension $\frac{d}{2}+ \nu + \frac{1}{2}$, while the fermionic double trace flow would give a fermionic operator of different dimension ($\frac{d}{2}+ \nu - \frac{1}{2}$).

The other possibility is to couple $\psi$ to $\mc O_{f+}$. For $1/2<\nu<1$, the IR limit of (\ref{eq:staticGpsi}) is, up to contact terms,
\be\label{eq:psiOp}
\langle \psi(p) \bar \psi(p)\rangle \approx i \sigma \cdot p\, \frac{c_{\nu-1}|g|^2(p^2-i\epsilon)^{\nu-1}}{m^2}\,,
\ee
from which we deduce that the dimension is $\Delta_{f+}$. This behavior is consistent with the fact that the interaction $g \psi \mc O_{f+}$ is relevant, but the induced double trace deformation is actually an irrelevant perturbation of the $g=0$ fixed point, so the dimension of $\mc O_{f+}$ is not modified in the IR. Finally, for $\nu >1$ both $\psi \mc O_f$ and $\mc O_f^2$ are irrelevant; in the IR $\psi$ decouples and $\mc O_+$ does not flow, as can be seen in (\ref{eq:staticGpsi}).

\subsection{Infrared dynamics of the spacetime dependent theory}

Let us now turn on a spacetime dependent interaction $g(x) \psi \mc O_{f+}$ that approaches a power law $g(x) = g_0 |x|^\alpha$ at long distances. A scaling argument near the $g=0$ fixed point gives $[g_0]=\alpha-(\nu-1)$, implying that time dependence makes this interaction relevant for $\alpha>\nu-1$. Similarly, the condition that the double trace $\frac{g^2}{m} \mc O_{f+}^2 \subset L$ be relevant is $\alpha>\nu - 1/2$.

The long distance correlators are computed as before using large-$N$ factorization and are described by the gaussian model of Appendix \S \ref{app:gaussian}. For large enough $\alpha$ as above, the self-energy for $\t \psi(x)= g(x) \psi(x)$ receives the dominant contribution from inverse of the $\mc O$ correlator (\ref{eq:Of2ptp}), giving
\be\label{eq:2-tpsi}
\langle \t \psi(p) \bar{\t \psi}(p) \rangle =i \sigma \cdot p\, \frac{1}{c_{\nu-1}(p^2-i\epsilon)^\nu}\,.
\ee
This is simply the previous static result (i.e.\ the limit of large $g$ in (\ref{eq:staticGpsi})) but now valid for $\nu>1$ as long as $\alpha>\nu - 1/2$. We note that this corresponds to a field of subunitary dimension $\Delta_{f-}$ but, as will be seen shortly, the propagator does not violate unitarity. Transforming back to position space and rewriting $\t \psi$ in terms of $\psi$ obtains
\be\label{eq:Gpsifinal}
\langle  \psi(x) \bar \psi(y) \rangle = - \frac{(2 \Delta_{f-}-1)}{c_{-\nu}c_{\nu-1}} \,\frac{i\sigma \cdot (x-y)}{g(x) [(x-y)^2]^{\frac{d}{2}-\nu+1}g(y)^*}\,.
\ee
This is our final form for the Green's function of $\psi$ in the regime where the effects from time dependence are dominant.

Eq.~(\ref{eq:Gpsifinal}) also allows us to determine the IR limit of the flow induced by a spacetime dependent double trace deformation (\ref{eq:doubletOf}) of the original CFT. The result is that at long distances the $\mc O_{f+}$ two-point function becomes
\be\label{eq:GOffinal}
\langle  \mc O_{f+}(x) \overline{\mc O}_{f+}(y) \rangle = - \frac{(2 \Delta_{f-}-1)}{c_{-\nu}c_{\nu-1}} \,\frac{i\sigma \cdot (x-y)m^2}{g(x)^{2} [(x-y)^2]^{\frac{d}{2}-\nu+1}g(y)^{*2}}\,.
\ee
As in the scalar case, we see that the power-law time dependent coupling leads to an (approximately) scale invariant regime with scaling dimensions modified by $\alpha$.

We should now discuss the unitarity constraints on this fermionic two-point function. For this, we couple $\t \psi$ to an external fermion $\chi$ by $\t \psi \chi + c.c. \subset L$ and demand that the $\chi \to \chi$ scattering amplitude obey the optical theorem~\cite{Grinstein:2008qk}. It is useful to recall how
this works in known static examples. Coupling $\chi \mc O_{f+}+ c.c. \subset L$, where the CFT fermionic operator $\mc O_{f+}$ has two-point function
\be
\langle \mc O_{f+}(p) \overline{\mc O}_{f+}(q) \rangle = i \sigma \cdot p\,c_{\nu-1} (p^2-i\epsilon)^{\nu-1}\,\delta^d(p-q)\,,
\ee
the scattering amplitude is
\be
\mc A(\chi \to \chi) \propto c_{\nu-1} (p^2-i\epsilon)^{\nu-1}\,.
\ee
The $i\epsilon$ prescription implies that in the forward lightcone ${\rm Im}\,\mc A \propto -c_{\nu-1} \sin(\pi (\nu-1))$, which is nonnegative for $\nu\ge0$. This reproduces the unitarity bound $\Delta_{f+} \geq (d-1)/2$.

Now we come to our time dependent theory, where the appearance of an operator of subunitary dimension $\Delta_{f-}$ in (\ref{eq:2-tpsi}) would naively suggest a violation of unitarity. However, this is avoided because the propagator also comes with a coefficient $c_{\nu-1}$. Coupling our probe fermion to $\tilde \psi$ and requiring ${\rm Im}\,\mc A\geq0$ obtains $\sin(\pi\nu)/c_{\nu-1}\geq0$. This is satisfied for all $\nu\geq0$. The conclusion is that as long as the UV dimension $\Delta_{f+}$ is above the unitarity bound, unitarity is preserved along the time dependent RG flow. 

\section{Unitarity bounds in supersymmetric gauge theories and FRW holography}\label{sec:SQCD}

So far we have worked with a simple class of large-$N$-solvable field theories exhibiting the effect of spacetime dependent couplings on infrared physics and unitarity.  We focused on the infrared, noting that in some cases, unitarity problems arise in the UV.  In those cases, the theory needs to be cut off or cross over to different physics at a sufficiently low scale to retain unitarity. This was implemented at the static level recently in \cite{Andrade:2011aa}, and arose in our more complicated FRW examples \cite{Dong:2011uf}.

In this section, we will combine the methods we developed in the previous section with those of supersymmetric gauge theory.
Of course, spacetime dependent couplings $g(x)$ break translation invariance and supersymmetry at a scale of order $\partial g/g$.  Nonetheless, this will prove useful for two reasons:

\noindent (i) Via the relation between dimensions and R-symmetries, it provides a beautiful method to relate unitarity bounds to the infrared dynamics of static theories (see e.g.~\cite{Seiberg:1994pq}).  We will be interested in comparing this to the infrared physics one obtains instead in the presence of spacetime dependent couplings.

\noindent (ii) Well above the scale $\partial g/g$, the results of the static theory pertain and as we will see can naturally provide a UV completion.

We begin by presenting a time-dependent version of an interesting class of models~\cite{Barnes:2004jj} with $\mc N=1$ supersymmetry; these have the structure of the above semiholographic models in the IR, and provide a concrete UV completion of that mechanism.  Then, we will return to our original motivation from~\cite{Dong:2011uf} and consider $\mc N=2$ supersymmetric gauge theories with time-independent couplings, such as those obtained on D3-branes probing parallel $(p,q)$ 7-branes.  In this class of theories, one can vary the hypermultiplet masses to obtain regimes where mutually nonlocal matter fields (electric and magnetic) become simultaneously light~\cite{magneticmatter}.  Our first question is why from a field theory point of view this class of theories never involves more than a certain number (here $n_*=12$) dyonic flavors descending from $(p,q)$ strings stretching between the D3-branes and the $(p,q)$ 7-branes.  We will trace this to a unitarity bound, which is evidently relaxed in the presence of appropriate time dependence as determined from the gravity side of the corresponding holographic models~\cite{Dong:2011uf}.

\subsection{$\mc N=1$ Supersymmetric QCD plus singlets:\ unitarity bounds and UV completion}
\label{subsec:SQCD}

In this section, we begin with the construction in \cite{Barnes:2004jj}.\footnote{We thank K. Intriligator for suggesting this class of examples to us.} In this class of models, in the deep UV one has an asymptotically free theory plus a singlet.  At sufficiently low energies (well below a dynamical scale $\Lambda$),~\cite{Barnes:2004jj} gives strong evidence that the theory consists of a nontrivial CFT with an irrelevant coupling to a (different) singlet $\Phi$, which hits the unitarity bound and decouples in the deep IR.  Consequently, if we consider a large-$N$ version of the theory then in this infrared regime the system boils down to a semi-holographic model of the kind we studied above.

Consider a four-dimensional $\mc N=1$ gauge theory with gauge group $SU(N_c)$, fundamental + antifundamental flavors $(Q, \t Q)$, and a set of singlets $S$ that couple to some of the flavors. The flavors are divided into two sets:\ $N_f'$ flavors $(Q_a', \t Q'_a)$ are coupled to $S$ by a superpotential
\be\label{eq:W1}
W = h S^{ab} {Q'}_a^\alpha {\t Q'}_{b\,\alpha}
\ee
(where $\alpha$ are the contracted gauge theory indices), while the remaining $N_f$ flavors $(Q_i, \t Q_i)$ have no superpotential interactions. For $h=0$ this is just SQCD with $N_f+N_f'$ flavors, which flows to a superconformal field theory in the window $\frac{3}{2} N_c < N_f + N_f' < 3N_c$. In this range, the superpotential (\ref{eq:W1}) is relevant, and drives the theory away from the $h=0$ fixed point.  In~\cite{Barnes:2004jj}, evidence was provided that the theory flows to another nontrivial SCFT.

Symmetries and anomaly cancellation are not enough to determine the dimensions of holomorphic operators at the putative new fixed point, but these dimensions can be found using $a$-maximization~\cite{Intriligator:2003jj, Barnes:2004jj}.
The result is that at the new fixed point where (\ref{eq:W1}) becomes marginal, the conformal dimensions are given by
\be\label{eq:dims}
\Delta(Q\t Q) = 2 y\;,\;\Delta(Q'\t Q') =  \frac{3(n+1-x) -2y}{n}\;,\;\Delta(S) = \frac{3(x-1)+2y}{n}
\ee
where
\be
x \equiv \frac{N_c}{N_f}\;,\;n \equiv \frac{N_f'}{N_f}
\ee
and the quantity $y$ (which is the prediction from $a$-maximization) is
\be\label{eq:ydef}
y = \frac{-3[2n(n+2)+(n(n-4)-1)x+x^2]+n\sqrt{9x^2(x-2n)^2-8n(n^2-1)x+4n^2}}{2x-2n(nx+4)}\,.
\ee
The gauge-invariant degrees of freedom are given by the mesons and baryons
\be\label{eq:singlets}
\Phi = (Q \t Q)\;,\;P = (Q \t Q')\;,\;P'=(Q' \t Q)\;,\;B_r= (Q^r Q'^{N_c-r})\,,
\ee
whose dimensions are determined from those in (\ref{eq:dims}).

The nature of the infrared theory changes when
\be
x > x_c(n) \equiv \frac{1}{3}+ \frac{5}{3}n - \frac{1}{3}\sqrt{1-14n+13n^2}\,.
\ee
If (\ref{eq:W1}) were still nonzero and marginal,
the meson $\Phi=(Q \t Q)$ would violate the unitarity bound:
\be
\Delta(\Phi) = 2 \Delta(Q) < 1\,.
\ee
Instead of breaking unitarity, what happens at this point is that the meson decouples, and for $x > x_c(n)$ we are left with a consistent CFT plus a free field $\Phi$.\footnote{$n \ge 2$ is required here, to avoid baryons hitting the unitarity bound. Eventually, for large enough $x$ the theory exits the conformal window and moves into the free magnetic phase.}

Since for $x \lesssim x_c(n)$ $\Phi$ is weakly coupled, we can write down an effective action
\be\label{eq:Leff}
L = L_{SCFT} + \int d^4 \theta\,\Phi^\dag \Phi + \int d^2 \theta\,\lambda\,\Phi \mc O + c.c.
\ee
with $\mc O$ being an operator in the CFT (which we identify below) of dimension
\be
\Delta(\mc O) = 3 - 2y\,.
\ee
For $x = x_c(n)$, 
$\Phi$ flows to a free field, implying that the $\Phi \mc O$ interaction is irrelevant so that $\Phi$ decouples. We see that this theory in the IR is like our semiholographic model. The main difference is that the gauge theory is unitary all the way to the UV; around the dynamical scale $\Lambda$ the composite nature of $\Phi$ emerges. (Also, in the static theory the dimension of $\Phi$ is calculated using supersymmetry, and no large-$N$ limit is required.)

In fact, Seiberg duality~\cite{Seiberg:1994pq} provides an explicit description for (\ref{eq:Leff}). Recall that the dual of $SU(N_c)$ SQCD with $N_F$ flavors $(Q, \t Q)$ has a gauge group $SU(N_F-N_c)$, $N_F$ magnetic quarks $(q, \t q)$ and $N_F^2$ singlets $M$ with a superpotential
\be
W_{mag}= q M \t q\,.
\ee
The singlets just correspond to the mesons $M = (Q \t Q)$.

The duality generalizes very simply to the theory with extra singlets as well. Start in the UV with $N_F=N_f+ N_f'$ flavors and dualize. At lower energy scales, below the dynamical scale the superpotential (\ref{eq:W1}) becomes a mass term $S^{ab} (Q' \t Q')_{ab} \subset W$ that lifts the singlets $S$ and the mesons $(Q' \t Q')$. Hence, at low energies the dual is a theory with gauge group $SU(N_f+N_f'-N_c)$, $N_f$ magnetic quarks $(q, \t q)$ (dual to $(Q, \t Q)$), $N_f'$ quarks $(q', \t q')$ (dual to $(Q, \t Q)$), and singlets $\Phi$, $P$, $P'$ (identified in (\ref{eq:singlets})) with superpotential
\be
W_{mag}= q \Phi \t q+ q P \t q'+ q' P' \t q\,.
\ee

In the magnetic theory, the meson $\Phi$ appears as an elementary field that couples to the rest of the fields via $\Phi q \t q \subset W$. So the operator $\mc O$ above is identified with the product of magnetic quarks $q \t q$. In this dual theory, when $x \ge x_c(n)$ the coupling $\Phi q \t q$ is irrelevant and the elementary $\Phi$ becomes a free field, in agreement with the analysis in the electric theory.

The dynamics of this system can be studied at large $N$. Indeed, the CFT exists at large $N_c$ and $N_f'$ with $N_f$ fixed, in which case loops containing $\Phi$ will be suppressed by powers of $N_f/N_c$ and $N_f/N_f'$.
Therefore in components we obtain a fermion $\psi$ and a complex scalar $\phi$ interacting with a large-$N$ CFT sector. The fermion couples linearly to the CFT operator $q \psi_{\t q}+  \t q\psi_q$ (where $\psi_q$ is the fermion component of the chiral superfield $q$ and so on). 
In this large-$N$ limit, the theory is in the same class of theories as our semiholographic models, and can be analyzed using similar techniques.

In particular, we wish to understand what effect a spacetime dependent coupling $\lambda(x)$ would have on the infrared physics.
Let us discuss this using the magnetic dual, which is more appropriate for this purpose.
We start with
\be\label{eq:Wdef}
W_{mag} \supset \lambda(x) q\Phi \t q\,,
\ee
where we continue to use supersymmmetric language to package the following component Lagrangian.
Denoting the scalar and fermion components of $\Phi$ by $\phi$ and $\psi$ respectively, the resulting classical component Lagrangian is
\begin{align}\label{eq:LSQCD}
L =& L_{SCFT} - \partial_\mu \phi^* \partial^\mu \phi- i \bar \psi \not \! \partial \psi - \lambda(x)^2 |\phi|^2 (|q|^2+ |\t q|^2) - \lambda(x)^2 |q \t q|^2 - \left(\lambda(x) \phi \psi_q \psi_{\t q} \right.\nonumber\\
& \left.-  \lambda(x) \psi (q \psi_{\t q}+  \t q\psi_q) +c.c.\right)
\end{align}
In the deep IR, supersymmetry no longer controls our calculations, but using large $N$ we can control the semiholographic mixing between $\psi$ and $q \psi_{\t q}+  \t q\psi_q$ in the same way as we did in \S \ref{sec:fermion}, obtaining correlators which include a would-be subunitary operator combined with factors of $1/\lambda$ which restore unitarity.  At shorter scales than $\lambda/\partial\lambda$, we return to the static theory, for which the double trace deformation generated by integrating out $\psi$ is irrelevant.  This theory is unitary, altogether providing a UV completion of our basic semi-holographic mechanism. In \S \ref{sec:fermion} we analyzed the fermionic case explicitly. It would also be interesting to study the full (\ref{eq:LSQCD}) at large $N$, including the effects of the quartic couplings which are absent in the theory analyzed in \S\S \ref{sec:semiholographic} and \ref{sec:fermion}.

\subsubsection{Further remarks}

Here we have considered a spacetime dependent coupling directly in the magnetic theory, which by itself provides a consistent UV completion for our semiholographic model. The electric theory gives a different UV theory (the two are equivalent only in the deep IR), and it is interesting to ask how to incorporate time dependence in this setup. One possibility would be to consider time dependent gauge couplings. Another option is to add new singlets $N$ and $\Phi$ to the electric theory, with superpotential
\be
W_{el} \supset Q N \t Q + \lambda(x) N \Phi\,.
\ee
This is the dual of (\ref{eq:Wdef}) when $\lambda$ is small. Note that for constant $\lambda$ the second term gives masses to $N$ and $\Phi$; integrating out $N$ then yields the usual relation between $\Phi$ and the electric meson, $(Q \t Q)= - \lambda \Phi$. In the time dependent case we should instead keep these fields in the effective theory; for $x \ge x_c(n)$ we expect that $\lambda(x)$ will modify the dynamics of $(Q \t Q)$, which would otherwise decouple in the static limit.

Finally, let us briefly mention another type of modification of the theory which could adjust the infrared physics in a similar way:\ add an extra chiral superfield $X$, assumed to couple in the superpotential as
\be\label{eq:invW}
W \sim \lambda \frac{q \Phi \t q}{X^k}
\ee
to good approximation at sufficiently large and homogenous $X$, with $k$ some positive number (and $\lambda$ constant). For instance, instanton-generated superpotentials can contain inverse powers of $X$. If there is a limit where we can treat $X$ as a fixed background, then requiring that (\ref{eq:invW}) is marginal at the fixed point obtains
\be
\Delta(\Phi) = 3 - \Delta(\mc O) + k \Delta(X)
\ee
with $\mc O = q \t q$ as before. So the dimension of $\Phi$ would be increased by its interaction with $X$, leading to unitarity consistent with nonvanishing $\lambda$, unlike the original static theory in which $\lambda\to 0$ in the infrared.  We have not constructed an example of this yet; it would be interesting to find a UV completion of this mechanism.

\subsection{Seiberg-Witten theory and flavor bounds}

Our next class of theories is an $\mc N=2$ gauge theory in four dimensions, studied by Seiberg and Witten~\cite{Seiberg:1994rs} as well as many other subsequent works\footnote{See e.g.~\cite{AlvarezGaume:1996mv} for a review and references.}.  Using these, we will make contact with our motivating examples \cite{Dong:2011uf}.  We will show that the matter content derived from the brane construction in \cite{Dong:2011uf}\ in the case of parallel 7-branes would not be unitary with static couplings.  This fits well with the fact that the dual description of the FRW spacetime solutions in \cite{Dong:2011uf}\ has time dependent couplings.  However, this theory is more complicated than those analyzed above; in particular we have not isolated a simple limit in which to calculate correlation functions on the field theory side.

A useful way to geometrize the Coulomb branch of $\mc N=2$ theories is as position collective coordinates of D3-branes probing sets of $(p,q)$ 7-branes.  In that language, we are interested in nontrivial fixed points which arise when a stack of $N_c$ color branes hits a set of mutually nonlocal 7-branes.  As the 3-branes approach the 7-branes, we obtain light mutually nonlocal flavors \cite{Ftheory, magneticmatter}.  We would like to use field theory techniques to understand how many such flavors can arise in this way.

Let us focus on the case of an SU(2) gauge group, with a one complex dimensional Coulomb branch labelled by $u$.  This captures the dynamics of the center of mass of the stack of D3-branes in the class of models just described.  We work near a fixed point which we take to sit at $u=0$.
Holomorphic quantities in the theory such as the gauge coupling function, the metric on the Coulomb branch, and BPS particle masses are related to the Seiberg-Witten (SW) curve
\be\label{eq:curve}
y^2 = x^3 -f(u) x -g(u)
\ee
where $x$ and $y$ are complex variables, $u$ is the local Coulomb branch coordinate and the singularity is located at $u=0$. $f$ and $g$ are polynomials in $u$; varying their coefficients corresponds to turning on relevant deformations of the fixed point. The SW curve defines a differential $\lambda_{SW}$ by
\be
\frac{d \lambda_{SW}}{du}= \frac{dx}{y}\,.
\ee
BPS masses are given by combinations of the periods of $\lambda_{SW}$ on the torus (\ref{eq:curve}); these periods are denoted by $(a,a_D)$ and are functions of $u$. Moreover, the scalar kinetic term is obtained by differentiating the K\"ahler potential
\be\label{eq:K}
K = \textrm{Im}(A_D \bar A)\,.
\ee

The Seiberg-Witten curve also determines the dimension of the Coulomb branch coordinate $u$~\cite{Argyres:1995xn}; let us review how this comes about. Since the BPS masses are linear combinations of $a$ and $a_D$, these periods have dimension one. This is also consistent with the K\"ahler potential above having dimension 2. Since $a \sim (u/y) dx$, we find the relation between the conformal dimensions
\be\label{eq:aconstr}
\Delta(x) - \Delta(y) + \Delta(u) =1\,.
\ee
Close enough to the singularity $u=0$ that describes the CFT, $f$ and $g$ have a power-law dependence, which we parametrize by
\be
f(u) \sim u^r\;,\;g(u) \sim u^s\,.
\ee
For $2s<3r$ ($2s>3r$) the singularity is dominated by $g$ (resp.~$f$), so these cases need to be treated separately.

First, when $2s<3r$, the SW curve for $u \to 0$ is $y^2 \approx x^3 - u^s$; since at the fixed point it should scale homogeneously under dilatations, the scaling dimensions obey
\be
2 \Delta(y) = 3 \Delta(x) = s \Delta(u)\,.
\ee
Combining this with (\ref{eq:aconstr}) we obtain
\be\label{eq:Delu1}
\Delta(u) = \frac{6}{6-s}\,.
\ee
We see that for $s>6$ the Coulomb branch field $u$ would violate unitarity. On the other hand, for $2s>3r$ the singularity is dominated by $f$, and a similar analysis leads to the dimension
\be\label{eq:Delu2}
\Delta(u) = \frac{4}{4-r}\,.
\ee
This violates unitarity for $r>4$. The set of all integers $(r,s)$ allowed by unitarity reproduce the ADE classification of singularities~\cite{Minahan:1996fg}.

Now we want to express $\Delta(u)$ in terms of the number $N_f$ of flavors -- i.e.\ the number of 7-branes in the $D3$--$(p,q)$7-brane realization of SW. $N_f$ is given by the order of the vanishing of the discriminant
\be
\Delta = 4f^3 +27 g^2
\ee
at the singularity. This relates $N_f$ to $(s,r)$ above:\ for $2s<3r$ we have $N_f=2s$, while $N_f=3r$ if $2s>3r$. In terms of $N_f$, both (\ref{eq:Delu1}) and (\ref{eq:Delu2}) reduce to
\be\label{eq:Delu}
\Delta(u) = \frac{12}{12-N_f}\,.
\ee
We conclude that there is no unitary $\mc N=2$ supersymmetric conformal field theory in this class of models for $N_f>12$.

At this point it is useful to make contact with AdS/CFT and review how this result is obtained in the gravity side. The F-theory description of $N_f$ 7-branes in Sen's limit of constant axio-dilaton gives a 10d metric
\be
ds^2 = \eta_{\mu\nu} dx^\mu dx^\nu + \frac{dz d\bar z}{(z \bar z)^{N_f/12}}\,.
\ee
Introducing a D3 probe parallel to the 7-branes, the worldvolume field $u$ that describes motion along the $z$ direction has a kinetic term
\be
S = -\int d^4 x \, \eta^{\mu\nu} \frac{\partial_\mu u \partial_\nu \bar u}{(u \bar u)^{N_f/12} }\,.
\ee
From here we can read off the scaling dimension $\Delta(u) = 12/(12-N_f)$, in agreement with the field theory result. (The relation between $u$ and $z$ that we used here is fixed by supersymmetry.) Nontrivial CFTs (some of which have no known field theory description) are obtained by placing $N_c$ D3-branes near one of these ADE singularities from 7-branes~\cite{Fayyazuddin:1998fb}.

The gravity side provides a useful description to explore mechanisms for avoiding the unitarity violation that we just found. In particular, controlled late-time time-dependent F-theory solutions with $N_f>12$ were found in~\cite{Kleban:2007kk} and the implications for the holographic duality were studied in~\cite{Dong:2011uf}.  There, we have a controlled gravity description with warping and hence a low energy field theory regime; but in that case the dual is strongly interacting and we have not done an independent field theoretic calculation of the effects of the time dependence.  From the gravity side we can see the dual theory is defined up to a time-dependent cutoff $\Lambda \sim 1/t$; it would be interesting to understand in more detail how this holographic cutoff is realized in the dual theory.  Gravity decouples in the dual theory at late times, so this question is purely field-theoretic.
Next, we will exhibit a static theory with $n>n_*$ {\it massive} flavors, another way to restore unitarity.

\subsection{Static theory with $n>n_*$ massive flavors}

In general, a field theory which is well defined in the UV (e.g.\ QED on the lattice above
its Landau pole) will do something sensible at all scales.  What we have shown is that it
cannot retain $\mc N=2$ SUSY and $n > n_*$  massless magnetic flavors.  In the discussion so far, we have focused on the effects of time dependence on unitarity conditions.  However, in this subsection we briefly mention another consistent flow in the static case in which $n>n_*$ magnetic flavors are present, but massive.

The work \cite{Polchinski:2009ch}\ constructed noncompact Calabi-Yau fourfolds describing intersecting $(p,q)$ 7-branes, generalizing those in \cite{Fayyazuddin:1998fb}.  These, combined with color $D3$-branes placed at the intersection, can produce AdS/CFT dual pairs with small internal dimensions.  In \cite{Polchinski:2009ch}\ a physical criterion for singularity-freedom was articulated, matching but generalizing standard results.  This limits how many 7-branes can intersect, leading to bounds of the sort described above (the details of which depend on the codimension of intersection).  If we take a case with $n>n_*$, i.e.\ with a singular intersection, and deform it so that the 7-branes do not intersect at the same point, this can remove the singularity.  It was argued in \cite{Fayyazuddin:1998fb}\ that generically for such F-theory configurations adding D3 color branes produces a good AdS/CFT duality.  The deformation away from the singularity means that the flavors obtained from (p,q) strings are not simultaneously light; they are generically massive.         


%
%
%

Although it would take us too far afield to describe in detail here, we have used gauge linear sigma model techniques to construct an example of a noncompact Calabi-Yau fourfold describing a configuration of 7-branes with $n>n_*$, deformed away from the would-be singular intersection.  In the resulting dual theory, the excessive magnetic flavors are massive, corresponding to a smoothed out tip controlled by a deformation parameter in the superpotential of the sigma model. The model has nontrivial RG flow corresponding to bending of the branes as one evolves in the radial direction toward the resolved singularity.

\section{Future directions}\label{sec:concl}

\noindent{\it Happy families are all alike; every unhappy family is unhappy in its own way. -- Leo Tolstoy.}
\smallskip

In this paper, we have shown how infrared physics and unitarity conditions are consistently modified in time dependent quantum field theory, as compared to static versions of the same theory.  We focused on double trace deformations and related semiholograpic theories in order to explicitly analyze these effects in tractable examples with nontrivial operator scaling.      
The lesson is more general, and suggests several interesting directions to pursue.  

We have seen how spacetime dependent couplings affect RG trajectories, in some cases reversing the direction of flow at long distances as compared to the static version of the theory.  Similarly, in cases which are classically marginal at the static level, spacetime dependent couplings will generically introduce nontrivial flow at the same order.  This opens up the possibility of new fixed points, for example in spacetime dependent QED in four dimensions with sufficiently many flavors to screen the interaction at one loop.    

Our results suggest a similar explanation for how the flavor content of the time dependent holographic quantum field theory in \cite{Dong:2011uf}\ (dual to an FRW geometry with a warped metric) is distinct from the flavor content allowed in the corresponding static theories.  It would be interesting to pursue other examples to see how unitarity bounds are affected by spacetime dependent couplings.  Examples to which we may apply these ideas include minimal models which are nonunitary at the static level, additional examples of SUSY and/or holographic gauge theories going below their static unitarity bound, and no-ghost theorems in various holographic backgrounds modified by time dependence. 

\section*{Acknowledgments}
We are grateful to T. Banks, L. Fitzpatrick, D. Harlow,  K. Intriligator, D. Jafferis, J. Kaplan, D. Marolf, M. Peskin, J. Polchinski, and S. Shenker for very useful discussions.  XD, BH, and ES thank the Kavli Institute for Theoretical Physics for hospitality during parts of this project.  This work is supported in part by the National Science Foundation under grant PHY05-51164, by the NSF under grant PHY-0756174, and by the Department of Energy under contract DE-AC03-76SF00515.

\section*{Appendix}
\appendix

\section{Dynamical couplings and unitarity}\label{app:dynamicalg}

In this section, we promote the time dependent coupling to a dynamical field, and study unitarity constraints and scales in this `parent' theory.  This will provide a useful partial UV completion of the main semiholographic example in the text, one which requires a cutoff that is static and well above the energy scale of the time variation of the effective coupling coming from the rolling scalar field.  A UV completion up to arbitrarily high scales is provided by the UV-supersymmetric examples in the main text.

The full action is
\be\label{eq:Lstatic}
L = L_{CFT}- \frac{1}{2}\left((\partial \phi)^2+ m^2 \phi^2 \right)+\lambda_0\, g \phi \mc O_+ - \frac{1}{2} (\partial g)^2 - V(g)
\ee
where we have promoted the coupling $g(x)$ to a field. Ignoring the cubic interaction $g \phi \mc O$, we will construct a potential $V(g)$ such that at late times $g(x) \approx g_0 t^\alpha$. Then we will determine the conditions under which $g(x)$ may be considered as a background field and we can neglect effects of the cubic interaction on the dynamics of this field.

The starting Lagrangian (\ref{eq:Lstatic}) has only static couplings, and it will be found that $V(g)$ does not involve higher dimension operators over the range of field of interest to us. However, at the UV fixed point the cubic coupling is irrelevant, $[\lambda_0]=2-\Delta_+<0$, for $d \ge 2$ and $\nu > 1$. Thus, this theory needs a UV cutoff
\be\label{eq:cutoff1}
\Lambda_0 = \frac{1}{\lambda_0^{1/(\Delta_+-2)}}\,.
\ee

Using conservation of energy,
\be
\frac{1}{2} \left(\frac{dg}{dt} \right)^2+V(g) = V_0
\ee
and inverting $t= (g/g_0)^{1/\alpha}$ gives a (late time) potential
\be\label{Vofg}
V(g) = V_0 - \frac{1}{2} g_0^{\frac{2}{\alpha}} \alpha^2 g(x)^{2({1- \frac{1}{\alpha}})}\,.
\ee
The potential is negative for large enough $g$, but the power $0< 2(1- \frac{1}{\alpha})< 2$ for $\alpha>0$. So  $g$ takes an infinite time to reach infinity and the system does not require a self-adjoint extension.

For our purposes, we are concerned about unitarity bounds but not model-building ``taste bounds"; i.e.\ we are content to tune the potential as required to maintain the shape (\ref{Vofg}).   Similarly, we may avoid strong effects of particle (or unparticle) production by coupling $\mathcal{O}$ to a sufficient number of additional operators in the CFT which do not couple directly to $g$.  Thus, as the time dependent motion of $g$ produces excited states of ${\cal O}$, it quickly shares that energy with other modes which do not directly backreact on $g$.  This procedure may buy us a parametrically long time under which nonadiabatic effects on the underlying state of the system may be ignored, though at some point these effects become important.

The energy scale for the perturbations of the rolling scalar is given by the square root of
\be
V''(g) = - (\alpha-1)(\alpha-2) \frac{g_0^{2/\alpha}}{g(x)^{2/\alpha}} = - (\alpha-1)(\alpha-2) \frac{1}{t^2}\,,
\ee
which is of the same order as $\partial g/g$.
The range $1< \alpha<2$ is interesting in that $V''(g)>0$, giving the fluctuation $\delta g$ a positive mass of order $1/t$.  Above the scale $\partial g/g$ we must include the dynamics of the field $g(x)$.  This gives a UV completion up to the scale $\Lambda_0$ (\ref{eq:cutoff1}) at which we could introduce a (static) cutoff to render the theory completely well defined.   As a result, the theory must be unitary on all scales; this is checked in the infrared in the main text.

\section{Gaussian theories with spacetime dependent masses}\label{app:gaussian}

In this appendix we consider a model of time dependence given by a quadratic action with two fields coupled together via spacetime dependent masses, and perform the path integral explicitly. The kinetic terms are taken to be arbitrary kernels $K(x,y)$. This theory also captures the leading result for the effective action of a field coupled to a large-$N$ CFT that was analyzed in \S \ref{sec:semiholographic} and \S \ref{sec:fermion}.

\subsection{Bosonic model}

Consider a Gaussian bosonic theory of the form
\be\label{eq:Sgaussian1}
S = - \frac{1}{2} \int d^dx d^d y\left(\phi(x) K_\phi(x,y) \phi(y) + \mc O(x) K_\mc{O}(x,y) \mc O(y)\right) + \int d^dx\,g(x) \mc O(x) \phi(x)\,.
\ee
In the setup of \S \ref{sec:semiholographic}, $\mc O$ would be an operator in a nontrivial CFT, $\phi$ is the weakly coupled scalar, and $g(x)$ is the spacetime dependent coupling. Also, in this case the kernels $K_\phi$ and $K_{\mc O}$ would be translationally invariant, but in our present discussion these kernels will be arbitrary. (Of course, they should be positive definite, so that the path integral is well defined.)

Our goal is to compute the two-point functions for (\ref{eq:Sgaussian1}). For this, it is convenient to adopt a matrix notation where
\be
S= - \frac{1}{2} \phi_a K_{ab} \phi_b
\ee
and
\be\label{eq:Kab}
K_{ab}= \left(\begin{matrix} K_\phi(x,y) && -g(x) \delta(x-y) \\ -g(x) \delta(x-y) && K_{\mc O}(x,y) \end{matrix} \right)
\ee
Here $a,b$ are multi-indices that distinguish both $(\phi, \mc O)$ and the position coordinates; summation over repeated indices also contains an integral over positions. With our mostly plus sign conventions, we have
\be
Z[J] = \int \mc D \phi\, \exp\left[- \frac{i}{2} \phi_a K_{ab} \phi_b + J_a \phi_a\right] = \exp\left[- \frac{i}{2} J_a (K^{-1})_{ab} J_b\right]
\ee
and the two-point function is given by
\be
\langle \phi_a \phi_b \rangle= \frac{\delta^2 Z}{\delta J_a \delta J_b} = -i (K^{-1})_{ab}\,.
\ee

The inverse of (\ref{eq:Kab}) can be calculated using the formula for the inverse of a block matrix
\be\label{eq:blockI}
\left(\begin{matrix}
A && B \\ C && D
\end{matrix} \right)^{-1} =
\left(
\begin{matrix}
(A-B D^{-1} C)^{-1} && - A^{-1} B (D- C A^{-1} B)^{-1} \\
-D^{-1} C (A - B D^{-1} C)^{-1} && (D - C A^{-1} B)^{-1}
\end{matrix}
\right)\,,
\ee
which gives
\begin{eqnarray}
(K^{-1})_{ab}&=&
\left(\begin{matrix}
(K_\phi - g K_{\mc O}^{-1}g)^{-1} &&  K_\phi^{-1} g (K_{\mc O}- g K_\phi^{-1}g)^{-1} \\ K_{\mc O}^{-1}g (K_\phi - g K_{\mc O}^{-1}g)^{-1}&& (K_{\mc O}- g K_\phi^{-1}g)^{-1}
\end{matrix} \right)\,\\
&=& 
\left(\begin{matrix}
(K_\phi - g K_{\mc O}^{-1}g)^{-1} &&  K_\phi^{-1} g (K_{\mc O}- g K_\phi^{-1}g)^{-1} \\ (K_{\mc O}- g K_\phi^{-1}g)^{-1} K_\phi^{-1} g && (K_{\mc O}- g K_\phi^{-1}g)^{-1}
\end{matrix} \right)\,
\end{eqnarray}
The two-point functions are then
\bea
\langle \phi(x) \phi(y) \rangle &=& - i(K_\phi - g K_{\mc O}^{-1}g)^{-1}(x,y) \nonumber\\
\langle \phi(x) \mc O(y) \rangle &=& - i\,\left[K_\phi^{-1} g (K_{\mc O}- g K_\phi^{-1}g)^{-1}\right](x,y) \nonumber\\
\langle \mc O(x) \mc O(y) \rangle &=& - i (K_{\mc O}- g K_\phi^{-1}g)^{-1}(x,y)\,.
\eea

These formulas reproduce the large-$N$ result (\ref{eq:phiphi}) for $\langle \phi(x) \phi(y) \rangle$ obtained by summing the geometric series with $\mc O$ propagators. However, it is important to note that for a kernel such as
\be
K_{\mc O}^{-1}(x,y)= \frac{i}{[(x-y)^2]^{\Delta}}
\ee
we need to add the rest of the CFT (that is responsible for the nontrivial dimension $\Delta$) in order to have a consistent theory. So, while the gaussian model captures correctly the two-point functions for $\phi$ and $\mc O$, one should keep in mind that it is not complete by itself without the remaining CFT. This is what we mean by `effectively gaussian' in the main text.

\subsection{Fermionic model}

Let us now analyze a fermionic Gaussian theory with spacetime dependent coupling $g$:
\be
S= \bar \psi K_\psi \psi - \frac{1}{2} m (\psi \psi + \bar \psi \bar \psi)+ \overline{\mc O}_f K_f \mc O + g \psi \mc O_f + \eta \psi + \eta_f \mc O_f + c.c.
\ee
where the sources $\eta$ and $\eta_f$ are introduced to calculate the propagators, and are set to zero at the end. For instance, $K_\psi = - i \bar \sigma^\mu \partial_\mu\,\delta^d(x-y)$ for a free fermion; we follow the two-component notation of~\cite{twocomp}. As before, $\mc O_f$ plays the role of the CFT fermionic operator, so we have not included a mass term for this field.

It is useful to introduce a `Majorana' fermion $\Psi_a \equiv (\psi, \bar \psi, \mc O_f, \overline{\mc O}_f)$, and similarly for the sources $ N_a \equiv (\bar \eta, \eta, \bar \eta_f, \eta_f)$, such that the action can be written compactly as
\be
S = \frac{1}{2}\,\bar \Psi_a K_{ab} \Psi_b + \bar \Psi_a N_a\,,
\ee
with
\be
K_{ab}= \left(
\begin{matrix}
K_\psi && -m && 0 && g^* \\
-m && \bar K_\psi && g && 0 \\
0 && g^* && K_f && 0 \\
g && 0 && 0 && \bar K_f
\end{matrix}
\right)\,.
\ee
Here $\bar K_\psi$ is defined by $\psi \bar K_\psi \bar \psi = \bar \psi K_\psi \psi$, and similarly for $\bar K_f$. In particular, $\bar K_\psi = - i \sigma^\mu \partial_\mu\,\delta^d(x-y)$ for a free fermion.

The equation of motion gives $K_{ab} \Psi_b=-N_a$, so the partition function becomes
\be
Z[N_a] = \int \mc D \Psi\, e^{iS}= \exp \left[-\frac{i}{2} \bar N_a (K^{-1})_{ab} N_b \right]\,.
\ee
From here, the two-point functions are
\be
\langle \Psi_a \bar \Psi_b \rangle = - \frac{\delta^2 Z}{\delta N_b \delta \bar N_a} = i (K^{-1})_{ab}\,.
\ee
The inverse $(K^{-1})_{ab}$ is calculated using (\ref{eq:blockI}). In the notation of this equation we have, for instance,
\be\label{eq:sub-blockpsi}
A - B D^{-1} C = \left(
\begin{matrix}
K_\psi - g^* \bar K_f^{-1} g && -m \\
-m && \bar K_\psi - g K_f^{-1} g^*
\end{matrix}
\right)\,.
\ee
This shows that quantum effects from $\mc O_f$ shift the effective action of $\psi$ by $K_\psi \to K_\psi - g^* \bar K_f^{-1} g$. The inverse of (\ref{eq:sub-blockpsi}) gives
\bea
\langle \psi(x) \bar \psi(y) \rangle &=& i \left[ (\bar K_\psi- g K_f^{-1}g^*) \cdot \left(( K_\psi- g^* \bar K_f^{-1}g) (\bar K_\psi- g K_f^{-1}g^*) -m^2\right)^{-1}\right](x,y) \nonumber\\
\langle \psi(x) \psi(y) \rangle &=& im \left(( K_\psi- g^* \bar K_f^{-1}g) (\bar K_\psi- g K_f^{-1}g^*) -m^2\right)^{-1}(x,y) \,,
\eea
in agreement with the large-$N$ results in \S \ref{sec:fermion}.

\section{Expansion of the two-point functions}\label{app:exp}

\subsection{Expansion around the nontrivial scale-invariant regime}\label{subsec:expn}

The two-point function \eqref{eq:finalGphi} suggests that there is an approximately scale-invariant regime in our semi-holographic model. In the following we show that our theory indeed flows to this regime in the IR for a range of $\alpha$.

For this we start with the full two-point function \eqref{eq:phiphi} and evaluate the corrections to \eqref{eq:finalGphi} from $K_\phi$. These are obtained by expanding (\ref{eq:phiphi}) around $K_\phi=0$:
\be\label{eq:pexpK}
\langle\phi(x)\phi(x')\rangle =i \left[(g K_{\mc O_+}^{-1} g)^{-1}+(g K_{\mc O_+}^{-1} g)^{-1} K_\phi (g K_{\mc O_+}^{-1} g)^{-1}+\cdots \right](x,x')\,.
\ee
The leading term here is given by (\ref{eq:finalGphi}). The subleading term from $K_\phi$ can be neglected if
\be
\left|(g K_{\mc O_+}^{-1} g)^{-1} K_\phi (g K_{\mc O_+}^{-1} g)^{-1}(x,x')\right|\ll \left|(g K_{\mc O_+}^{-1} g)^{-1}(x,x')\right|\,,
\ee
or more directly
\be\label{eq:corr}
\left|[(x-x')^2]^{\Delta_-} \int d^dz\, \frac{1}{g(z)[(x-z)^2]^{\Delta_-}}\,(-\partial_z^2 + m^2) \frac{1}{g(z)[(z-x')^2]^{\Delta_-}}\right| \ll 1\,.
\ee
Let us estimate this integral in euclidean space for the power law coupling $g(x)=g_0|x|^\alpha$. Convergence at the origin (in the Euclidean version of this calculation) requires $\alpha<(d-2)/2$ when $\phi$ has a kinetic term, or $\alpha<d/2$ if it only has a mass term, but as mentioned above we need not extend the power law form of $g(x)$ all the way to the origin. Convergence at $z \rightarrow x, x'$ requires $\nu > 0, 1$ respectively, however, these divergences are present in the static case as well and can be cancelled by local counterterms.  Convergence at large $|z|$ requires $\alpha>2\nu-d/2$ in the massive case $m\ne0$, or $\alpha>2\nu-d/2-1$ in the massless case $m=0$.

We would like to show \eqref{eq:corr} for large $|x|$, $|x'|$, and $|x-x'|$. There are (at least) two ways of taking this limit: either $|x|\sim|x'|\sim|x-x'|$, or one of them is much larger -- say $|x|\gg|x'|$. In either case, a leading contribution to the integral comes from the region $|z|\sim|x'|$ and the left hand side of \eqref{eq:corr} is of order
\be\label{eq:kpcr}
\sim \left(-\frac{1}{|x'|^2}+m^2 \right) \frac{1}{|x'|^{2(\alpha-\nu)}}\,.
\ee
One may verify that higher order corrections in \eqref{eq:pexpK} all come with additional powers of \eqref{eq:kpcr}. Therefore all corrections from $K_\phi$ are negligible for large $|x|$, $|x'|$ if we have
\be\label{eq:relm}
\alpha>\nu
\ee
in the massive case $m\ne0$, or
\be\label{eq:relmz}
\alpha>\nu-1
\ee
in the massless case $m=0$. These conditions are strictly stronger than the corresponding convergence conditions at large $|z|$, as long as $\Delta_-=d/2-\nu>0$. We conclude that our theory flows to an IR fixed point characterized by the two-point function \eqref{eq:finalGphi} when \eqref{eq:relm} or \eqref{eq:relmz} is satisfied.

The analog calculation in Minkowski signature would have additional divergences on the light cone if $\Delta_->1$. These divergences are present already in the static limit, and are absent given an appropriate $i\epsilon$ prescription which is easy to implement in momentum space.  In our case as well, this 
is simplest to address by Fourier transforming the factors $1/[(x-z)^2]^{\Delta_-}$ and $1/[(x'-z)^2]^{\Delta_-}$, then integrating over $z$.  This turns the left hand side of (\ref{eq:corr}) into an expression proportional to (considering for simplicity the term proportional to $m^2$)
\beq\label{eq:pcorr}
m^2c_\nu^2\left|[(x-x')^2]^{\Delta_-}\int d^d p_1\int d^d p_2 \tilde{\left(\frac{1}{g^2}\right)}(p_1+p_2) \frac{e^{i(p_1 x+p_2 x')}}{(p_1^2-i\epsilon)^\nu(p_2^2-i\epsilon)^\nu}\right|
\eeq
with $\tilde{\left(\frac{1}{g^2}\right)}(p_1+p_2)$ the Fourier Transform of $1/g^2(z)$, a smooth function of $z$.
For even $\alpha$, this can also be done by analytically continuing the (regular) Euclidean result. 

It is also worth noting that the scale covariance manifest in our IR two-point functions does not hold for all times, as we have explained above.
In particular, our theory has (un)particle production due to the spacetime-dependent coupling, and $1/N$ corrections also become important at sufficiently late times.

\subsection{Expansion around the free fixed point}\label{subsec:expf}

We may also ask if there are other IR fixed points to which our theory could flow. Specifically, we may ask if the free $g=0$ fixed point characterized by \eqref{eq:kokp} is IR stable.

Let us first work in the massless case $m=0$, so the fixed point contains a free massless scalar field decoupled from a CFT. To investigate its stability in the IR, we expand the full two-point function \eqref{eq:phiphi} around $g=0$:
\be\label{eq:pexpg}
\langle\phi(x)\phi(x')\rangle =-i \left(K_\phi^{-1}+K_\phi^{-1} g K_{\mc O_+}^{-1} g K_\phi^{-1}+\cdots\right)(x,x')\,.
\ee
The leading term here is $1/|x-x'|^{d-2}$, the two-point function of a free massless scalar field. The subleading term can be neglected if
\be\label{eq:fmzc}
\left||x-x'|^{d-2}\int d^d z\int d^d z'\frac{g(z)g(z')}{|x-z|^{d-2}|z-z'|^{2\Delta_+}|x'-z'|^{d-2}}\right|\ll 1\,.
\ee
Again one could consider different ways of taking $|x|$, $|x'|$, and $|x-x'|$ to large values. Convergence of the integral at large $|z|$, $|z'|$ leads to a weaker condition than our final result. The left hand side of \eqref{eq:fmzc} is estimated to be of order $|x'|^{2(\alpha-\nu+1)}$. Therefore we conclude that the free fixed point is IR stable when $\alpha<\nu-1$. When this condition is violated, our previous result shows that the theory flows to the nontrivial fixed point characterized by \eqref{eq:finalGphi}.

Note that we would have arrived at the same conclusion (\ref{eq:relmz}) had we demanded that the $g(x)g(x')$ term in the effective action be a relevant deformation of the free fixed point. Indeed, under the scaling transformation
\be
x \to \lambda x\;,\;x' \to \lambda x'\;,\;\phi \to \lambda^{-\frac{d-2}{2}} \phi
\ee
the last term in (\ref{eq:Seff}) is relevant precisely when $\alpha> \nu-1$. So our previous explicit calculation shows that we are in a situation where scaling arguments from static QFT still apply in the presence of spacetime dependent couplings. The power law $g(x)=g_0|x|^\alpha$ has turned an interaction that would have been irrelevant in the static case into a relevant one (for $\alpha> \nu-1$). The scale at which this coupling crosses over from irrelevant to relevant is $\Delta p \sim 1/|x|$. In other words, this is a ``dangerously irrelevant'' operator, with the ``danger'' coming from spacetime dependence.

One could also consider the massive case $m\ne0$, where the scalar field is gapped out from the free fixed point in the IR. In this case we look at the two-point function of $\mc O_+$.\footnote{One should in principle analyze $\langle\mc O_+\mc O_+\rangle$ also at the nontrivial scale-invariant regime, but there we have an approximate operator equation $\mc O_+(x)=m^2\phi/g(x)$ in the IR, so the analysis is similar. In the massless case we have a different operator equation $\mc O_+(x)=-\partial^2\phi(x)/g(x)$, so $\mc O_+$ is the descendant of $\phi$ up to a factor of $g(x)$.} In the large-$N$ limit it similarly sums to
\be\label{eq:oo}
\langle\mc O_+(x)\mc O_+(x')\rangle=-i\left(K_{\mc O_+}-gK_{\phi}^{-1}g\right)^{-1}(x,x')\,.
\ee
The free fixed point is stable in the IR if we can ignore the subleading corrections in the expansion
\be
\langle\mc O_+(x)\mc O_+(x')\rangle=-i\left(K_{\mc O_+}^{-1}-K_{\mc O_+}^{-1}gK_\phi^{-1}gK_{\mc O_+}^{-1}+\cdots\right)(x,x')\,,
\ee
where the propagator $K_\phi^{-1}(x,y)$ is approximately a delta function $\delta^d(x-y)/m^2$ in the IR. Therefore we arrive at the following condition
\be
\left||x-x'|^{2\Delta_+}\int d^d z\int d^d z'\frac{\delta^d(z-z')g(z)g(z')}{m^2|x-z|^{2\Delta_+}|x'-z'|^{2\Delta_+}}\right|\ll 1\,,
\ee
which can be verified to be equivalent to $\alpha<\nu$ for large $|x|$, $|x|'$, and $|x-x'|$. Again, we conclude that our theory flows to the free fixed point in the IR if $\alpha<\nu$, otherwise it flows to the nontrivial scale-invariant regime characterized by \eqref{eq:finalGphi}.

\section{Higher-derivative toy model}\label{app:higherderiv}

In this paper we have focused on models that are static in the UV and exhibit time dependence at energies below the scale $\partial g/g$.  It is also possible that the scale $\partial g/g$ lies above the scale $\Lambda_g$ (a scale at which in the static theory ghosts can appear).  With time dependent couplings, the role of $\Lambda_g$ is different from what it is in the static theory; the would-be ghost solutions must be reanalyzed.  Here we set up that question in a simple toy model.    

Consider a scalar field
\beq\label{higherderiv}
S=\int d^d x \left\{(\partial\phi)^2 + g(t)(\partial^2\phi)^2+\dots \right\}
\eeq
There are additional solutions to the wave equation since the action is fourth order rather than second order.  These solutions represent a breakdown of the theory in the standard quantization\footnote{It is however possible to quantize the time-independent theory under an alternate quantization such that the spectrum is ghost-free, since the Hamiltonian is PT-symmetric.  For a discussion of these issues, see e.g.\ \cite{ghostfreefourthorder}.}, however, as long as the theory is cut off in the UV before the scale at which the new solutions enter it will be well-defined and unitary.
With time dependence, there will be additional terms involving $\dot{g}, \ddot{g}$ in the equation of motion which may remove the new solutions over some range of scales.  We can analyze this explicitly in a class of models with $g(t) = g_0 t^{\alpha}$.

The equations of motion following from the action are
\beq\label{freeeom}
-\nabla^2\phi+\nabla^2((g(t)\nabla^2\phi) = 0.
\eeq
It is clear from this that there is always a solution $\nabla^2\phi=0$.  For constant $g(t)$, there is additionally a solution with $\nabla^2\phi \sim \phi/g^2$.  However, for $\frac{\dot\phi\dot g}{\phi g}$ or $\frac{\ddot g}{g} \gg 1/g$ this solution is absent, the ansatz fails.  The toy example makes the origin of the $O(\frac{1}{t})$ cutoff (which we found also in the semiholographic example) particularly transparent: for $\ddot{g} \gg 1,$ the overall scale $g_0$ of the coupling cancels from the equation of motion, and the solutions depend on the scales of the derivatives $\frac{\partial g}{g},$ $\left(\frac{\partial^2 g}{g}\right)^{1/2}.$

Working for simplicity in flat space, for the time-independent case $\alpha = 0,$ the solutions are of the form $\phi(t, \vec{x}) \propto e^{i \omega t + i \vec{k}\vec{x}},$ where
\beq
\omega = \pm |\vec{k}|, \pm \sqrt{\frac{1}{g} - \vec{k}^2}.
\eeq
\noindent The first pair of solutions corresponds to the `normal' solutions with $\partial^2 \phi = 0$.


For general $\alpha \neq 1,$ in the regime $\ddot{g} \sim \frac{g_0}{t^2} \gg 1$ we can drop the first term in \ref{freeeom}.  The solutions are separable in space and time and can be found explicitly.  For a spatial ansatz $e^{i \vec{k}\vec{x}},$ there are a pair of normal solutions with time-dependent factor $e^{\pm i k t}$ and an additional pair whose positive frequency solution is
\beq
\propto e^{i k t}\left(\frac{-2 i k t^\alpha}{1-\alpha}\right) + e^{-i k t} e^{\frac{-i \pi \alpha}{2}} (2 k)^\alpha \Gamma\lbrack 1-\alpha, -2 i k t \rbrack
\eeq
\noindent where we have made use of the incomplete gamma function.  The additional (positive frequency) solution in the case $\alpha = 1$ is
\beq
\propto e^{i k t}\log(t)+e^{-i k t} Ei(2 i k t)
\eeq
The power-law dependence means that these solutions enter at an energy scale of $O(\frac{1}{t})$, which is above the static cutoff defined by the instantaneous value of the coupling for a range of parameters.  Thus the analysis of unitarity conditions will be significantly different in the time dependent version of this theory as compared to the static case.



\bibliographystyle{JHEP}
\renewcommand{\refname}{Bibliography}
\addcontentsline{toc}{section}{Bibliography}
\providecommand{\href}[2]{#2}\begingroup\raggedright

\end{document}